\documentclass[a4paper,11pt]{article}
\usepackage{jinstpub} 

\usepackage{multirow}%
\usepackage{booktabs}%
\usepackage{subcaption}
\usepackage{bm}
\usepackage{tikz}
\usetikzlibrary{arrows.meta}

\title{\boldmath Exploring ESS$\nu$SB Near Water Cherenkov Detector Designs Through Graph Neural Network Flavour Identification}

\author[a]{J.~Aguilar}
\author[b]{M.~Anastasopoulos}
\author[r]{D.~Barčot}
\author[c]{E.~Baussan}
\author[b]{A.K.~Bhattacharyya}
\author[b]{A.~Bignami}
\author[d,e]{M.~Blennow}
\author[f]{M.~Bogomilov}
\author[b]{B.~Bolling}
\author[c]{E.~Bouquerel}
% ...

\affiliation[a]{Consorcio ESS-bilbao, Parque Científico y Tecnológico de Bizkaia, Laida Bidea, Edificio 207-B, 48160 Derio, Bizkaia, Spain}
\affiliation[b]{European Spallation Source, Box 176, SE-221 00 Lund, Sweden}
\affiliation[c]{IPHC, Universit\'{e} de Strasbourg, CNRS/IN2P3, Strasbourg, France}
\affiliation[d]{Department of Physics, School of Engineering Sciences, KTH Royal Institute of Technology, 106 91 Stockholm, Sweden}
\affiliation[e]{The Oskar Klein Centre, AlbaNova University Center, Roslagstullsbacken 21, 106 91 Stockholm, Sweden}
\affiliation[f]{Sofia University St. Kliment Ohridski, Faculty of Physics, 1164 Sofia, Bulgaria}
% ...
\affiliation[r]{Center of Excellence for Advanced Materials and Sensing Devices, Ruđer Bo\v{s}kovi\'c Institute, 10000 Zagreb, Croatia}
\affiliation[s]{Dipartimento di Matematica e Fisica, Universit\'a di Roma Tre, Via della Vasca Navale 84, 00146 Rome, Italy}
% ...

\emailAdd{alexander.burgman@fysik.su.se}
\emailAdd{joakim.cederkall@fysik.lu.se}
\emailAdd{peter.christiansen@fysik.lu.se}
\emailAdd{kaare.iversen@fysik.lu.se}
\emailAdd{jcpark@ibs.re.kr}

\abstract{The ESS$\nu$SB experiment aims to measure CP violation in the leptonic sector with high precision, necessitating robust reconstruction of neutrino events in the water Cherenkov (WC) detectors. In this work, we investigate the flavour identification potential of the proposed near WC detector using graph neural network (GNN)-based classification, with a focus on variations of key detector design parameters. In particular, we study whether a smaller and/or less instrumented detector can achieve the required classification performance.

Using detailed Monte Carlo simulations of charged-current (CC) neutrino interactions, we train GNN classifiers to distinguish electron and muon neutrino CC events. We find that GNN-based classification remains accurate even for detector configurations with volumes up to a factor of eight smaller than the nominal design, with only moderate degradation in classification efficiency at fixed background rejection. The resulting loss in efficiency can largely be compensated by increased exposure time. Furthermore, we demonstrate that reduced photomultiplier tube (PMT) coverage in the nominal detector has a limited impact on classification performance, provided that coverage is maintained in regions of highest signal yield, in particular near the forward end-cap.}


\begin{document}
\maketitle
\flushbottom

\section{Introduction} \label{sec:intro}

Understanding whether charge–parity (CP) symmetry is violated in the leptonic sector remains one of the central open questions in contemporary particle physics. Neutrino oscillations provide a unique avenue for probing this possibility, as the oscillation probabilities for neutrinos and antineutrinos depend explicitly on the CP‑violating phase $\delta_{\mathrm{CP}}$ \cite{Cervera2000}. Sensitivity to this parameter is enhanced at the second oscillation maximum, where the relative impact of $\delta_{\mathrm{CP}}$ on the appearance probability is significantly larger than at the first maximum \cite{Coloma2012}. However, the second maximum occurs at roughly three times the baseline distance of the first, and the associated flux reduction scales with the inverse square of the distance, necessitating nearly an order of magnitude higher luminosity to exploit this increased sensitivity.

One way to achieve the required event rates is to employ a neutrino source of unprecedented intensity. The 5 MW European Spallation Source (ESS) linear accelerator (linac) in Lund, Sweden \cite{Garoby2018}, can be extended to deliver a high-power neutrino beam. Building on this capability, the ESS$\nu$SB project has been proposed as a next-generation experiment optimized for measurement of leptonic CP violation \cite{Alekou2022, Alekou2023}. ESS$\nu$SB will compare the flavour composition of the beam at a near detector with that observed at a far detector located 340 km from the source, a distance chosen to sample both the first and second oscillation maxima for $\nu_\mu \rightarrow \nu_e$ and $\bar\nu_\mu \rightarrow \bar\nu_e$ transitions.

A prerequisite for achieving the physics goals of ESS$\nu$SB is the ability to reconstruct neutrino interactions with high accuracy and minimal bias, which is influenced both by the detector instrumentation and the reconstruction methods. It has been discussed how the necessary precision can be achieved with a 750 t near water Cherenkov detector and a 538 kt far detector using traditional likelihood-based reconstruction methods as proposed in the ESS$\nu$SB Conceptual Design Report \cite{Alekou2022, Burgman2022}. Here we want to explore modifications to the near detector design, to assert the potential of using a smaller and/or less instrumented detector as a means to reduce costs, while preserving the required flavour identification performance. Previous studies have discussed how Graph Neural Networks (GNNs) are well suited for classification of neutrino detector events, with good performance across a wide range of event topologies, with the additional advantage of being faster to train and run than traditional likelihood-based methods \cite{Abbasi2022, GNNKaare}. Classification with Machine Learning (ML) models such as GNNs is typically performed by applying a threshold to the model output score, which provides a natural definition of ambiguous events, corresponding to scores close to 0.5, rather than near 0 (high-confidence background) or near 1 (high-confidence signal). The smaller detector sizes explored in this work will result in fewer events where the produced charged lepton is both created and comes to rest inside the detector volume, which we refer to as fully contained events. We hope to show that with decreased detector size, the GNN model can learn to identify the events that are fully contained and ignore the events where the lepton is not contained by assigning them a score in the ambiguous range. Additionally, we expect the GNN to perform slightly better on these than traditional methods, as it has been shown to more easily handle low-energy events or events with pions \cite{GNNKaare}. Classification performance using GNNs is expected to be correlated with GNN performance for other reconstruction tasks necessary for the full analysis (such as vertex, energy, and direction regression), and the results obtained here can guide future full reconstruction studies, where the same techniques can be applied. 

The ESS$\nu$SB near‑detector complex, which will measure the flavour composition of the unoscillated neutrino beam, combines three complementary detection technologies:

\begin{itemize}
    \item a magnetized scintillator tracker (SFGD),
    \item an emulsion‑based detector similar to the NINJA system used at J‑PARC \cite{Fukuda2017}, and
    \item a water Cherenkov (WC) detector.
\end{itemize}

The WC detector, which is the focus of this work, is proposed to consist of a horizontal cylindrical tank measuring 10.96 m in length and 9.44 m in diameter, holding approximately 750 $\text{m}^3$ of ultra‑pure water \cite{Alekou2022}. Its interior will be instrumented with more than 22,000 photomultiplier tubes (PMTs) of 3.5‑inch diameter, providing about 30 \% optical coverage. This work will demonstrate how changes to these parameters impact GNN flavour classification performance.

\section{Methods}

\subsection{Simulation}\label{sec:simulation}

The investigations described in the following sections rely on the use of Monte Carlo simulations of neutrino interactions in varying configurations of the ESS$\nu$SB near WC detector. The neutrino interaction vertex generator GENIE v3.0.6 \cite{Andreopoulos2010, andreopoulos2015genie} was used to simulate the neutrino interactions with the water molecules in the detector tank, and the water Cherenkov detector simulation software WCSim \cite{WCSim} was used to simulate particle transport and the detector response. 

For this study, the neutrino interaction simulations created for the ESS$\nu$SB CDR was repurposed without modifications. For these simulations, muon neutrino and electron neutrinos with a uniform distribution of energies from 0 GeV to 1.5 GeV, were simulated with pure water as the target material. For more details, see \cite{Alekou2022}. To constrain the study, only charged current (CC) interactions were included. For an analysis of the performance on neutral current events with the nominal detector design, see \cite{GNNKaare}, which demonstrates how the NC events can be rejected using GNN model and thus do not affect performance. The accuracy of this result is expected to follow the same detector configuration dependence observed in the flavour classification studies presented in the following sections. Table \ref{tab:simulations} shows the distribution of simulated neutrino/antineutrino events by flavour.

\begin{table}
    \centering
    \begin{tabular}{cccc}
        & Simulated & Training (33\% validation) & Testing \\
        \hline
        \textbf{Neutrinos (CC)} & \textbf{400,000} & \textbf{100,000} & \textbf{300,000} \\
        $\nu_\mu$       & 100,000 & 25.000 & 75.000 \\
        $\bar \nu_\mu$  & 100,000 & 25.000 & 75.000 \\
        $\nu_e$         & 100,000 & 25.000 & 75.000 \\
        $\bar \nu_e$    & 100,000 & 25.000 & 75.000 \\
        \hline
    \end{tabular}
    \caption{Distribution of the training and test samples for the CC neutrino interaction data. The training was split into random samples of 67\% training data and 33\% validation data, thus only approximately maintaining the same ratios between flavours.}
    \label{tab:simulations}
\end{table}

\begin{figure}
    \centering
    \begin{tikzpicture}[scale=4.4]
        \draw [fill=gray, fill opacity=.25]
        (90:2.36mm) coordinate (a)
        -- ++(-5.48mm, 0) coordinate (b) 
        arc (90:270:1.75mm and 2.36mm) coordinate (d) 
        -- (d -| a) coordinate (c) arc (270:90:-1.75mm and 2.36mm); 
        \draw [fill=gray, fill opacity=.25]
        (0,0) coordinate (t) circle (1.75mm and 2.36mm);
        \draw [densely dashed] (d) arc (270:90:-1.75mm and 2.36mm);
        \draw []
        (90:2.975mm) coordinate (A)
        -- ++(-6.95mm, 0) coordinate (B) 
        arc (90:270:2.20mm and 2.975mm) coordinate (D) node [midway, right, anchor=west, inner sep=5pt] {$3$}
        -- (D -| A) coordinate (C) arc (270:90:-2.20mm and 2.975mm);
        \draw []
        (0,0mm) coordinate (T) circle (-2.20mm and 2.975mm);
        \draw [densely dashed, thin, opacity=.2] (D) arc (270:90:-2.20mm and 2.975mm);
        \draw []
        (90:3.745mm) coordinate (A)
        -- ++(-8.70mm,0) coordinate (B) 
        arc (90:270:2.78mm and 3.745mm) coordinate (D) node [midway, right, anchor=west, inner sep=5pt] {$2$}
        -- (D -| A) coordinate (C) arc (270:90:-2.78mm and 3.745mm);
        \draw []
        (0,0mm) coordinate (T) circle (-2.78mm and 3.745mm);
        \draw [densely dashed, thin, opacity=.3] (D) arc (270:90:-2.78mm and 3.745mm);
        \draw []
        (90:4.72mm) coordinate (A)
        -- ++(-10.96mm, 0) coordinate (B) 
        arc (90:270:3.5mm and 4.72mm) coordinate (D) node [midway, right, anchor=west, inner sep=5pt] {$1$}
        (D) node [midway, right, anchor=west, outer sep=108pt] {$4$}
        -- (D -| A) coordinate (C) arc (270:90:-3.5mm and 4.72mm);
        \draw []
        (0,0mm) coordinate (T) circle (-3.5mm and 4.72mm); 
        \draw [densely dashed, thin, opacity=.4] (D) arc (270:90:-3.5mm and 4.72mm);
        \draw[{Latex[scale=1.5]}-]
        (280:6mm)
        -- ++(-6.0mm,0) coordinate (S);
        \draw[]
        (S) node [anchor=east, inner sep=5pt] {Beam direction};
    \end{tikzpicture}
    \caption{Illustration of the detector dimensions used for the neutrino event simulations, to scale. \textbf{1:} nominal volume, \textbf{2:} half volume, \textbf{3:} 1/4 volume, and \textbf{4:} 1/8 volume.}
    \label{fig:detectorsizetikz}
\end{figure}
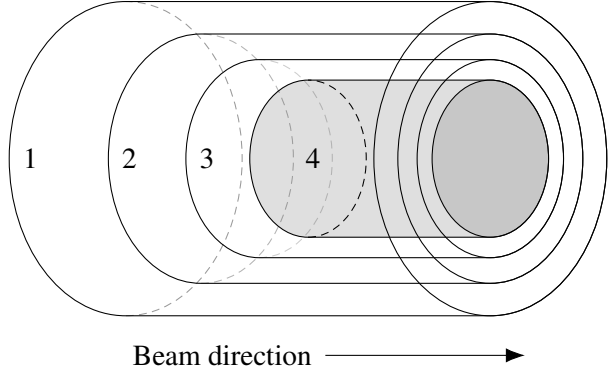

The simulated neutrino interactions were used to produce four datasets of detector responses, varying the size of the ESS$\nu$SB near detector. The nominal dataset uses dimensions identical to those selected and studied in the ESS$\nu$SB Conceptual Design Report \cite{Alekou2022}. The second and third datasets have the diameter and length both reduced by a factor $\sqrt[3]{2}$ and $\sqrt[3]{2}^2$, respectively, resulting in half and 1/4 the volume, and the fourth uses half the diameter and length of the baseline leading to 1/8 of the volume of the nominal size. All simulations were done using 4 inch PMTs of the Box\&Line type developed for Hyper-Kamiokande with 8.4 kHz dark noise ratio \cite{Tashiro:2021Og}. Table \ref{tab:detectorsizes} shows the four sets of detector sizes and coverage used for the response simulations, and Figure \ref{fig:detectorsizetikz} shows the geometry of the 4 detector sizes to scale for visual comparison. Variations in the PMT coverage was achieved after performing the detector response simulations as described in section \ref{sec:coverage} to avoid performing extra simulations.

\begin{table}
    \centering
    \begin{tabular}{ccccc}  
        & Length & Diameter & Volume & PMT Coverage \\
        \hline
        Nominal  & 10.96 m & 9.44 m & 767.1 m$^3$ & 30 \% \\
        Half volume & 8.70 m & 7.49 m & 383.5 m$^3$ & 30 \% \\
        1/4 volume & 6.95 m & 5.95 m & 191.8 m$^3$ & 30 \% \\
        1/8 volume & 5.48 m & 4.72 m & 95.9 m$^3$ & 30 \% \\
        \hline
    \end{tabular}
    \caption{Dimensions and coverage of the ESS$\nu$SB near detector tank for the four simulated datasets. PMT coverage was varied after simulating the datasets as described in Section \ref{sec:coverage}.}
    \label{tab:detectorsizes}
\end{table}

\subsection{GNN Flavour Classification}

Machine-learning–driven classification and reconstruction techniques have become increasingly important in high energy physics experiments, and specifically for neutrino experiments with Cherenkov-based detectors. Early applications have relied on convolutional neural networks (CNNs), which demonstrated strong performance on tasks such as particle identification and energy estimation \cite{Abbasi2021, Aiello_2020}. More recently, graph neural networks (GNNs) have emerged as a powerful alternative for Cherenkov detector event reconstruction \cite{Abbasi2022, Zhu2025}, and have been shown to be a good fit for ESS$\nu$SB detector events \cite{GNNKaare}. GNNs operate on graph-structured data, where events are represented as collections of nodes connected by edges, and information is exchanged between neighbouring nodes through iterative message-passing layers \cite{GNNs}. Conceptually, GNNs and CNNs are related in that both incorporate locality: they aggregate information from nearby elements in a chosen feature space. However, GNNs do so without requiring the underlying data to lie on a regular grid.

This flexibility is particularly relevant for Cherenkov detectors, where sensor positions are irregularly distributed and the geometry is inherently sparse. Applying CNNs in such settings typically requires mapping detector signals onto a fixed grid, introducing padding or interpolation steps that may distort spatial relationships or lead to unnecessary computation. In contrast, GNNs operate directly on the native detector topology. Although message passing can be computationally demanding, especially for graphs with $\mathcal{O}(10^3)$ nodes, where operations are performed per connected pair, the required network depth is often smaller than that of deep CNN architectures, mitigating total cost.

In this work, each neutrino interaction is represented as a single graph, where the nodes correspond to the PMT hits associated with the event. The node feature vector includes the PMT coordinates (x, y, z), the hit time t, and the recorded charge. For an event with $N$ PMT signals, edges are constructed using a K‑nearest‑neighbour procedure in the ($x$,$y$,$z$,$t$) space, yielding an input feature matrix $\mathbf{X} \in \mathbb{R}^{N \times 5}$ and an edge-index matrix $\mathbf{E} \in \mathbb{N}^{2 \times M}$, where $M=K \cdot N$ is the number of edges. All input features are scaled by fixed constants using the values from \cite{GNNKaare}, so that their mean absolute values are approximately 1, to work optimally with the GNN loss and activation functions.

Training and inference are performed using the GraphNeT framework \cite{Sogaard2023}, originally developed by the IceCube Collaboration \cite{Abbasi2012, Aartsen2017}. GraphNeT provides tools for constructing GNN-friendly graph representations from PMT-level data, efficient data loading and batching strategies, and interfaces to PyTorch \cite{paszke2019pytorchimperativestylehighperformance}, PyTorch Geometric \cite{FeyLenssen2019}, and PyTorch Lightning \cite{williamfalcon20203828935}. Following \cite{GNNKaare}, this study uses the DynEdge architecture, a one-cycle learning-rate scheduler \cite{smith2018disciplinedapproachneuralnetwork}, and early stopping applied to avoid overfitting. 

For each graph representing one event, the trained GNN outputs a scalar value between 0 and 1, interpreted as the predicted probability that the event corresponds to the signal flavour. Throughout this work, electron flavour is treated as the default signal class, meaning that true electron-neutrino events take value 1, unless otherwise specified.



\subsection{Coverage Adjustment}\label{sec:coverage}

In addition to studying the classification performance with varying detector dimensions, this study also aims to gauge the effects of variations in PMT coverage on the classification performance, and specifically the idea of using non-uniform coverage across the detector volume. Neutrino telescopes such as IceCube \cite{Aartsen2017}, or multi-purpose detectors such as Hyper-Kamiokande \cite{hyperK}, need to perform well for neutrinos coming from all directions, but for the near detector of an accelerator experiment like ESS$\nu$SB knowing that all the neutrinos of interest will come from the same direction means that some areas of the detector may benefit more from higher coverage than others.

\begin{figure}[!t]
    \centering
    \includegraphics[width=1\linewidth]{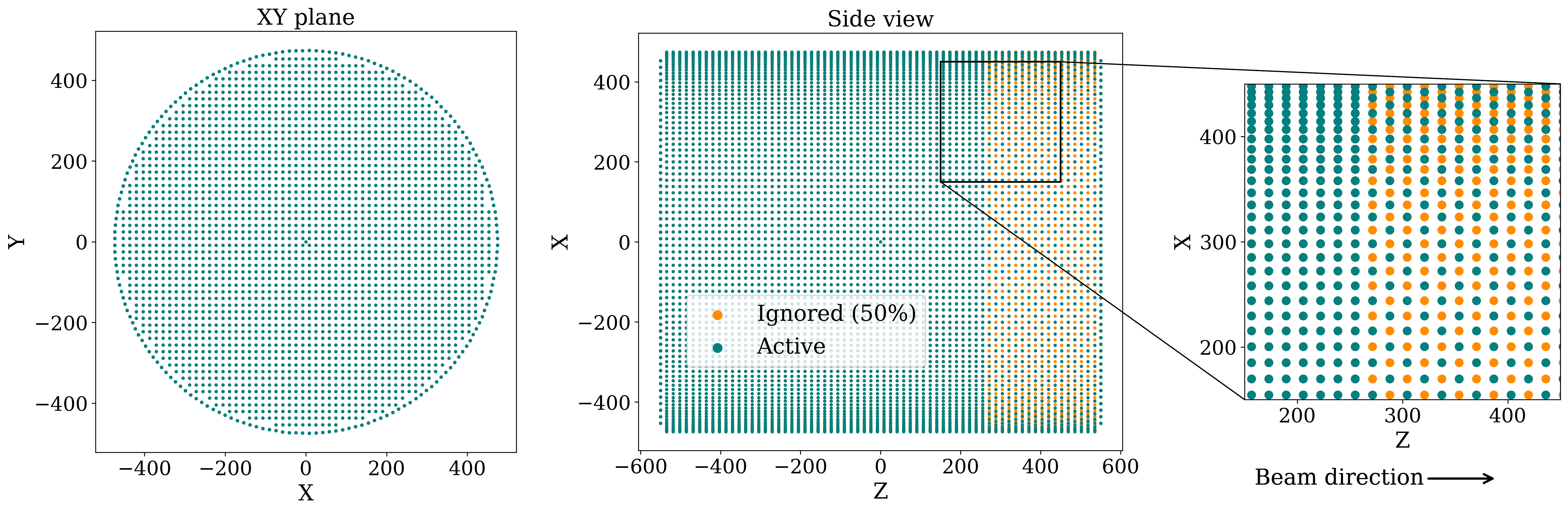}
    \caption{Visualisation of adjustment of the PMT coverage in the last quarter of the detector barrel (in the beam direction), Segment 4, by 50 \%. Hits from the ignored PMTs shown in orange will not contribute to the graphs made for one of the datasets created for study in the following sections.}
    \label{fig:PMTpos}
\end{figure}

While WCSim provides automatic generation of detector layouts according to the desired coverage, performing new simulations for each PMT layout to be studied can be computationally expensive. Instead, this study relies on using a single simulated dataset, and varying the coverage by simply ignoring the contributions of certain PMTs for each iteration. For instance, if one wanted to study the effects of 50 \% reduced coverage in the last quarter of the detector barrel (in the beam direction) as demonstrated in Figure \ref{fig:PMTpos}, PMT signals from every other PMT in this region would be ignored. In practice, this was done by creating a copy of the simulated dataset, discarding hits based on their (x, y, z) coordinates according to the desired pattern. While this requires additional storage space for each configuration, it is computationally advantageous over filtering the signals during training, which requires loading in unnecessary information for each event when building the graphs. 

\begin{figure}[!t]
    \centering
    \includegraphics[width=.5\linewidth, trim={3cm 2cm 0cm 5cm}, clip]{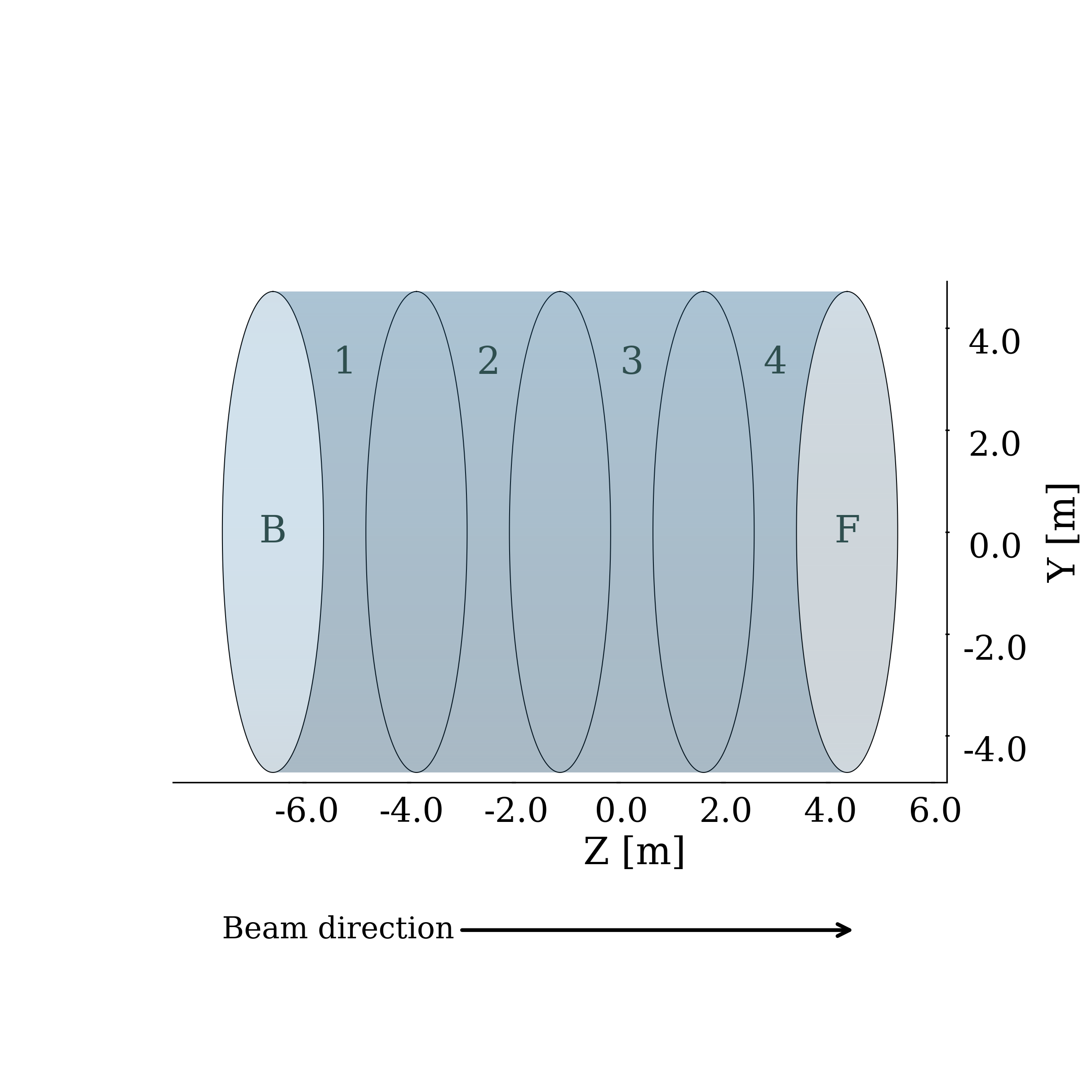}
    \caption{Illustration of the detector with indications of the 6 segments of PMTs for which the coverage will be varied in the following sections. Segments 1-4 are each one fourth of the barrel, Segment B is the backward end-cap in the beam direction and Segment F the forward end-cap.}
    \label{fig:segmentlabels}
\end{figure}

For this study, a simple scheme was chosen to illustrate the potential of this method. The detector PMT positions were divided into 6 segments, for which the PMT coverage was adjusted individually: 4 slices in the xy-plane of equal width of the barrel denoted Segments 1-4, and one each for the backward (Segment B) and forward (Segment F) end-caps. The scheme is illustrated in Figure \ref{fig:segmentlabels}, for easy comparison with the segments labels in plots in the later sections. This approach is flexible and can easily be extended to higher granularity or more complex schemes, without needing additional simulations.

\section{Results}

\subsection{The Importance of Retraining}

\begin{figure}[!t]
    \centering
    \includegraphics[width=1\linewidth]{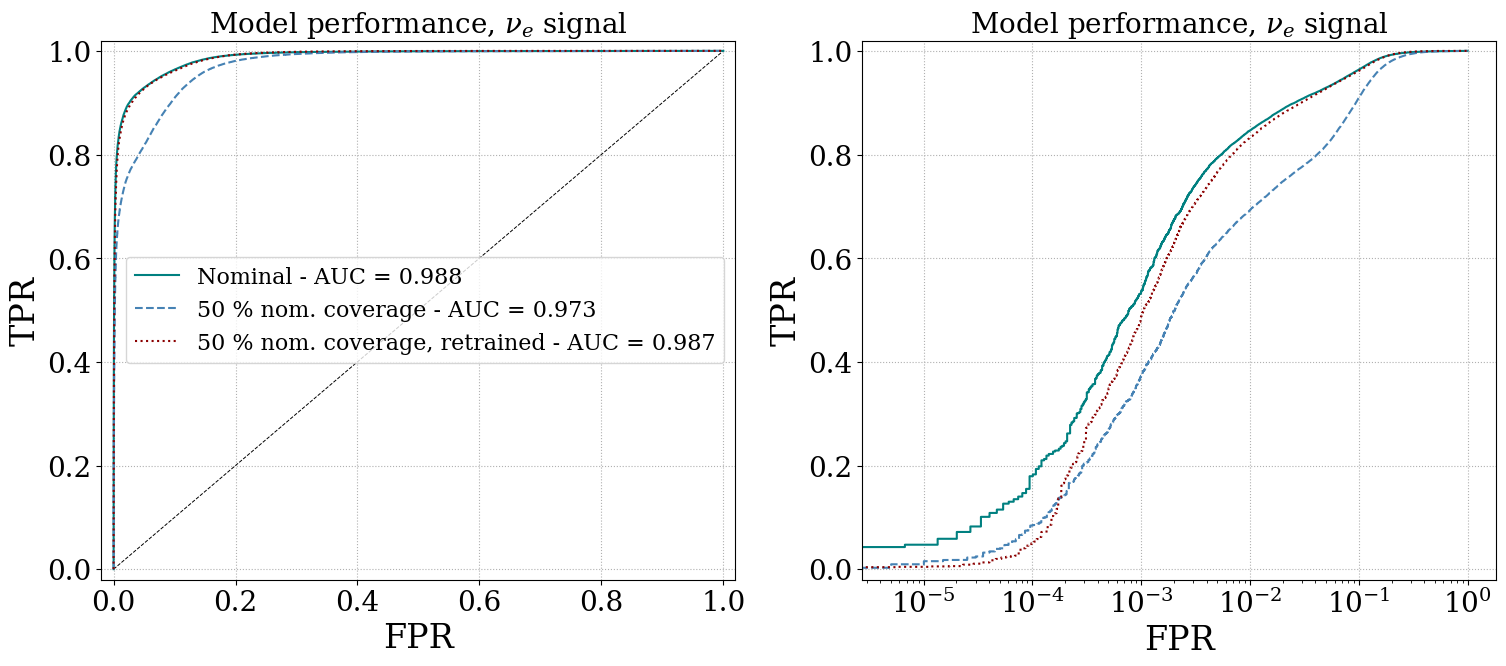}
    \caption{Receiver Operating Characteristic (ROC) curves for neutrino flavour classification using a baseline model (solid green), the same model on simulations ignoring 50 \% of the PMTs uniformly (dashed blue), and a second model trained with only 50 \% of the PMTs active (dotted red). The right shows the same figure with logarithmic y-axis to provide more detail in the $10^{-3}$ low false positive rate (FPR) region, which is the electron neutrino sample target.}
    \label{fig:trainingcompare}
\end{figure}
As a prerequisite for the following studies, the performance of using an already trained model on simulations made with a different configuration was studied. Figure \ref{fig:trainingcompare} shows Receiver Operating Characteristic (ROC) curves of flavour classification using a model trained on simulations for the nominal detector, with using all PMTs in green, and the same model used for prediction ignoring 50 \% of the PMTs in dashed blue, with a logarithmic x-axis on the right to visualize details in the low false positive rate (FPR) regions. The dotted red curve shows a new model trained on the same simulations, ignoring 50 \% of the PMTs.

It is clear from both visual inspection and by comparing the calculated area under the ROC curves (AUC) that there is a significantly greater decrease in performance from the baseline (AUC=0.988) without retraining (0.973) than with retraining (0.987). The retrained model has almost as good performance as the baseline, with significant difference only in the low FPR region where the resulting TPR is slightly lower.

Overall, this illustrates how much of the performance loss when using a less granular detector is due to lack of generalization of the GNN model compared to the lack of information. The larger difference in performance for the model without retraining, indicates that the performance loss is dominated by effects of lack of generalization. Thus, although it can be preferable to have a generalizable model and not need to retrain, the rest of this study will focus on retrained models to illustrate only the effects of loss of information when modifying the detector. Generalization is of course still important: In deployment, any selected model should be tested to be robust against detector configuration divergences that we may experience in real data-taking (such as inactive PMTs or lower signal yield).

\subsection{Detector Size Variation}\label{sec:detectorsize}

\begin{figure}[!t]
    \centering
    \includegraphics[width=1\linewidth]{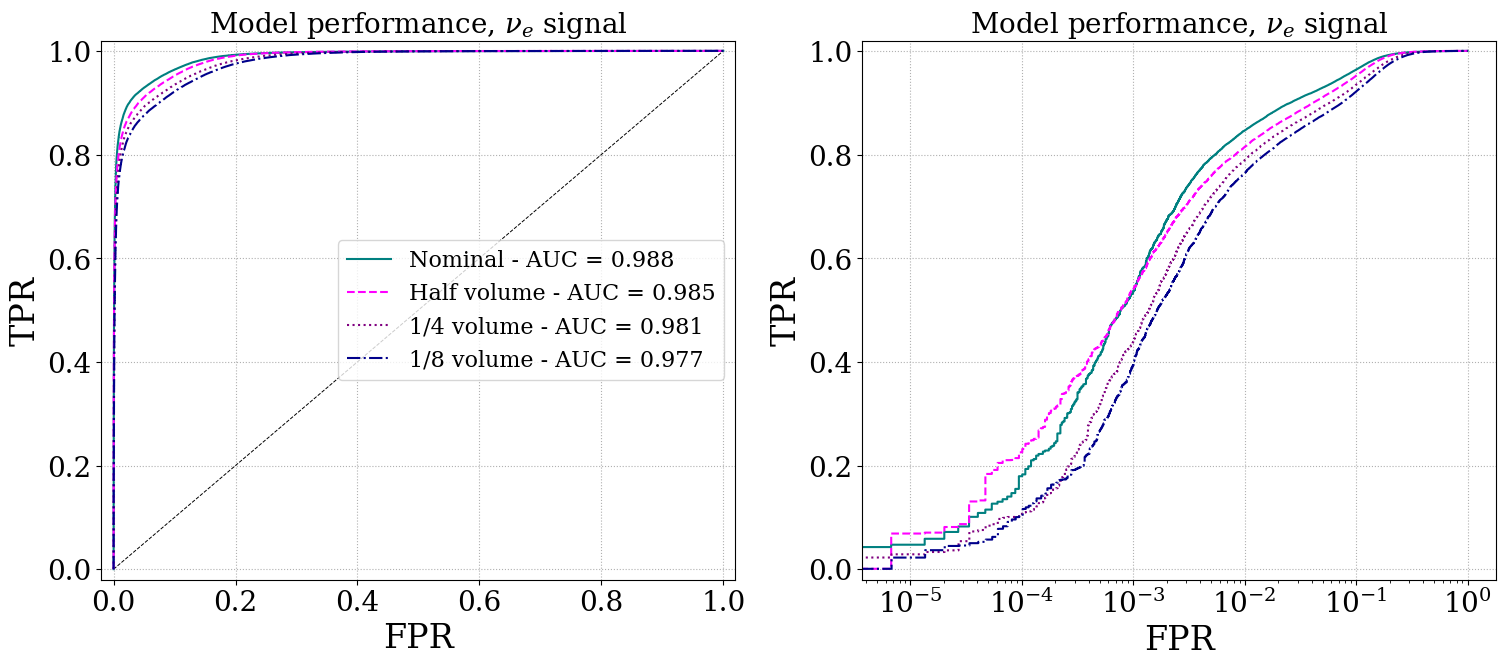}
    \caption{ROC curves for neutrino flavour classification models trained on the nominal, half volume, 1/4 volume, and 1/8 volume datasets, using electron neutrino flavour as signal. The figure on the right is the same as on the left but with logarithmic x-axis to highlight the performance in the $10^{-3}$ FPR region, which is the target FPR for the electron neutrino sample. The models trained on the datasets with smaller detector sizes perform overall worse than the baseline, with AUCs of 0.985, 0.981, and 0.977 for the half, 1/4, and 1/8 volume models, respectively, and at the $10^{-3}$ FPR, the resulting TPR is 53.5 \% for the nominal size, and similar at 53.9 \% for the half volume geometry, and decreases to 43.8 \% and 39.3 \% for the 1/4 and 1/8 volume geometries, respectively.}
    \label{fig:sizecompare}
\end{figure}

\begin{figure}[!t]
    \centering
    \includegraphics[width=1\linewidth]{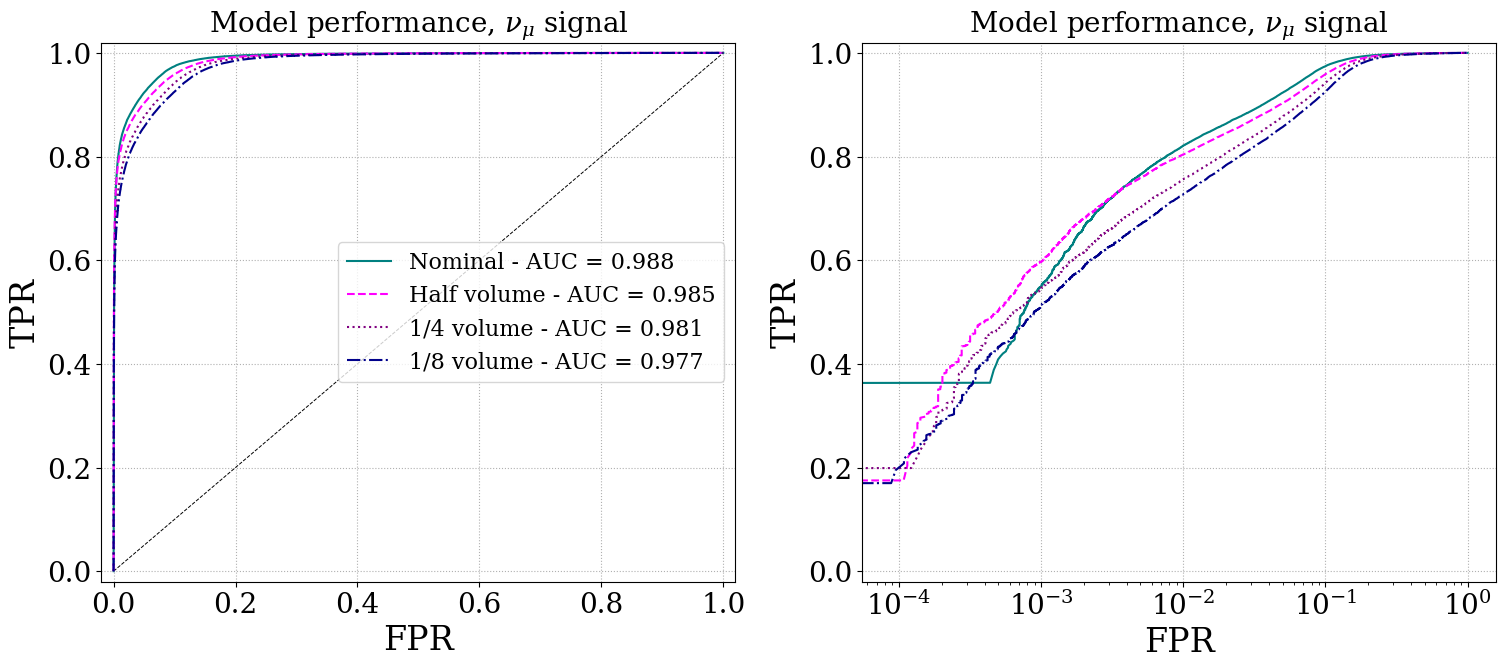}
    \caption{ROC curves for neutrino flavour classification models trained on the nominal, half volume, 1/4 volume, and 1/8 volume datasets, using muon neutrino flavour as signal. The figure on the right is the same as on the left but with logarithmic x-axis to highlight the performance in the $10^{-2}$ FPR region, which is the target FPR for the muon neutrino sample. The AUCs are by definition the same as with electron neutrino flavour as signal, but the at $10^{-2}$ FPR, the resulting TPR only decreases from 82.1 \% to 80.4 \%, 75.6 \%, and 72.7 \% for the half, 1/4, and 1/8 volume geometries, respectively.}
    \label{fig:sizecompare_background}
\end{figure}

To study the the potential of using the GNN approach to reconstruct events with reduced detector sizes, classification models were trained on the nominal, half volume, 1/4 volume, and 1/8 volume simulated datasets described in Section \ref{sec:simulation}. Figure \ref{fig:sizecompare} shows ROC curves for the four models, with a logarithmic x-axis on the right to provide more detail in the $10^{-3}$ FPR region. As expected, the overall classification performance decreases with detector size, with the smallest geometry (1/8 volume) obtaining the smallest AUC of 0.9779. However, the ROC AUC only gives a general measure of the model performance and the potential trade-off between TPR and FPR. At the desired FPR level ($10^{-3}$ for the electron neutrino sample), the resulting TPR is only reduced from 53.5 \% to 43.8 \% and 39.4 \% for the 1/4 volume and 1/8 volume geometries, respectively, while for the half volume, the FPR similar to the nominal at 53.9 \%. Figure \ref{fig:sizecompare_background} shows the same ROC curves, but using muon neutrino flavour as signal, and shows a similar trend with TPRs at the target $10^{-2}$ FPR decreasing with detector size, but remaining on the similar scale as the nominal detector. Rather than breaking down and being unable to reconstruct any events properly with a detector with half or even $1/8$ of the proposed volume, the model is able to separate a significant sample of electron neutrinos from the muon neutrino background. The reduction in TPR, or efficiency, is simply as loss of statistics which can be translated directly to increased runtime of the experiment to obtain the same results. As a result, these numbers illustrate the trade-off between detector size and runtime, which can be used to inform the design of the detector. This makes it attractive to consider smaller geometries for the Near Detector where the neutrino rate is high.

\begin{figure}[!t]
    \centering
    \includegraphics[width=.4\linewidth]{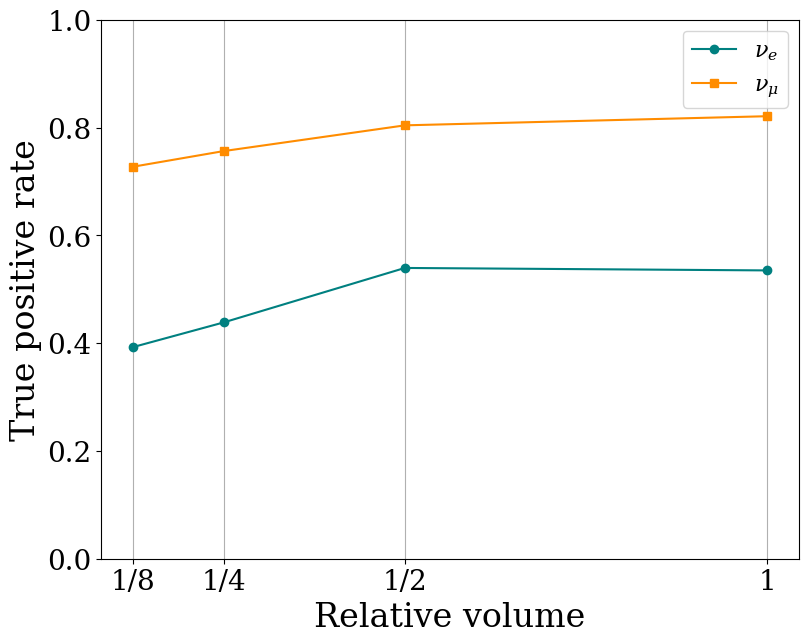}
    \caption{The TPRs obtained for the electron and neutrino event samples when adjusting the model score threshold to $10^{-2}$ and $10^{-3}$ FPR criteria, for each of the 4 detector simulations and a function of the detector size relative to the nominal size. In all cases the obtained TPR decreases with decreasing detector size, except the case for electron neutrino events with the 1/2 volume detector which shows no significant change from the nominal detector.}
    \label{fig:TPR_by_size}
\end{figure}

To summarize the behaviour, Figure \ref{fig:TPR_by_size} shows the relationship between detector volume and obtained TPR for electron and muon neutrinos for each of the four detector simulations. Here it is quite apparent that the change from nominal to half volume results in very little change, but the decrease in TPR is more significant for the 1/4 and 1/8 volumes for both flavours.

\begin{figure}[!t]
    \centering
    \includegraphics[width=1\linewidth]{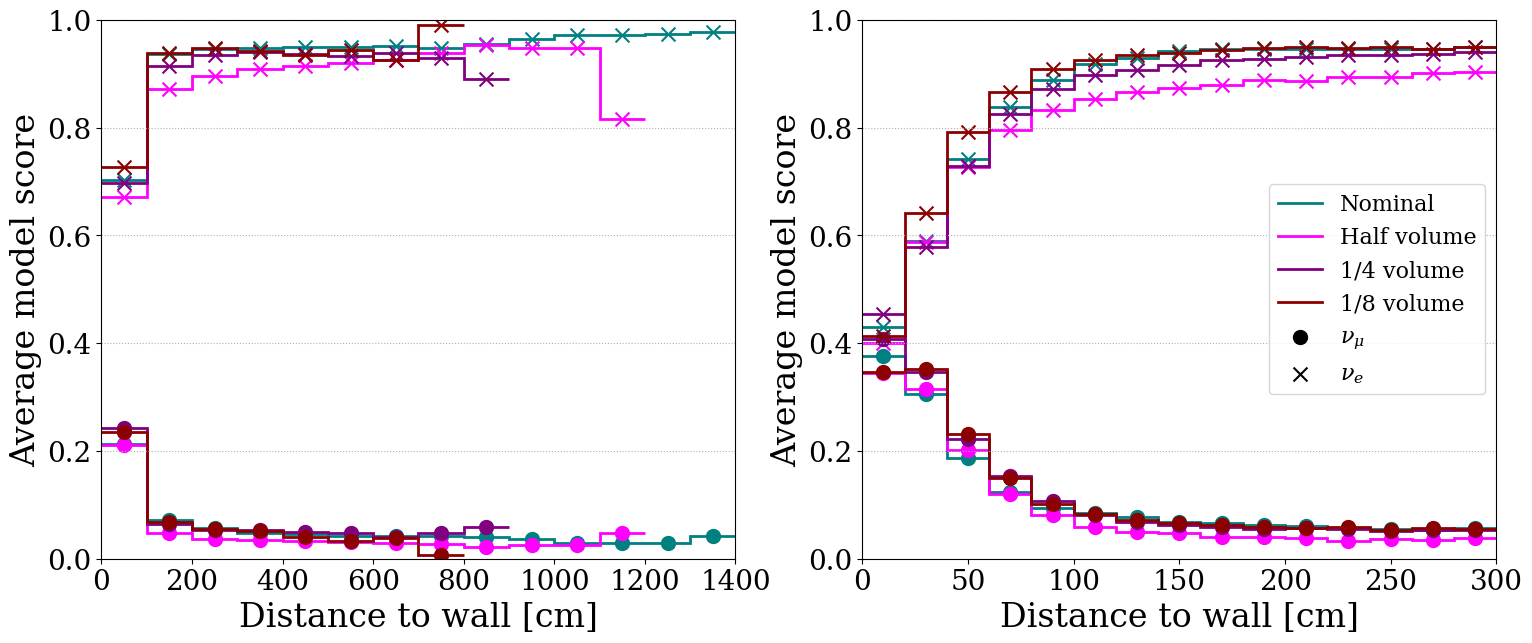}
    \caption{Average model score as a function of the distance from the interaction vertex to the detector wall along the lepton trajectory, for the nominal detector volume (green), the half volume (magenta), 1/4 volume (purple), and 1/8 volume (blue), for muon neutrinos (circles) and electron neutrinos (crosses). On the right, the same plot but zoomed to the 0-150 cm region. For all detector sizes, events nearest the wall are on average more ambiguous, receiving scores closer to 0.5, with scores converging towards the target scores of 0 and 1 for muon and electron neutrinos as distance increases. This matches the expectation that events that are not fully contained in the detector and thus leave less signal are harder to classify.}
    \label{fig:sizecompare_dist}
\end{figure}

To investigate the effect of the detector size on classification performance, the average model score for the nominal detector volume, the half, 1/4, and 1/8 volumes is plotted in Figure \ref{fig:sizecompare_dist} as a function of the distance from the interaction vertex to the detector wall along the lepton trajectory. In line with our expectations, the bins with the shortest distance to the wall (0-150 cm) have average model scores closer to 0.5, signifying more ambiguous events. As events where the lepton leaves the detector after travelling a short distance are expected to produce less Cherenkov light and less signal, these events are expected to be harder to classify on average. 

\begin{figure}[!t]
    \centering
    \includegraphics[width=1\linewidth]{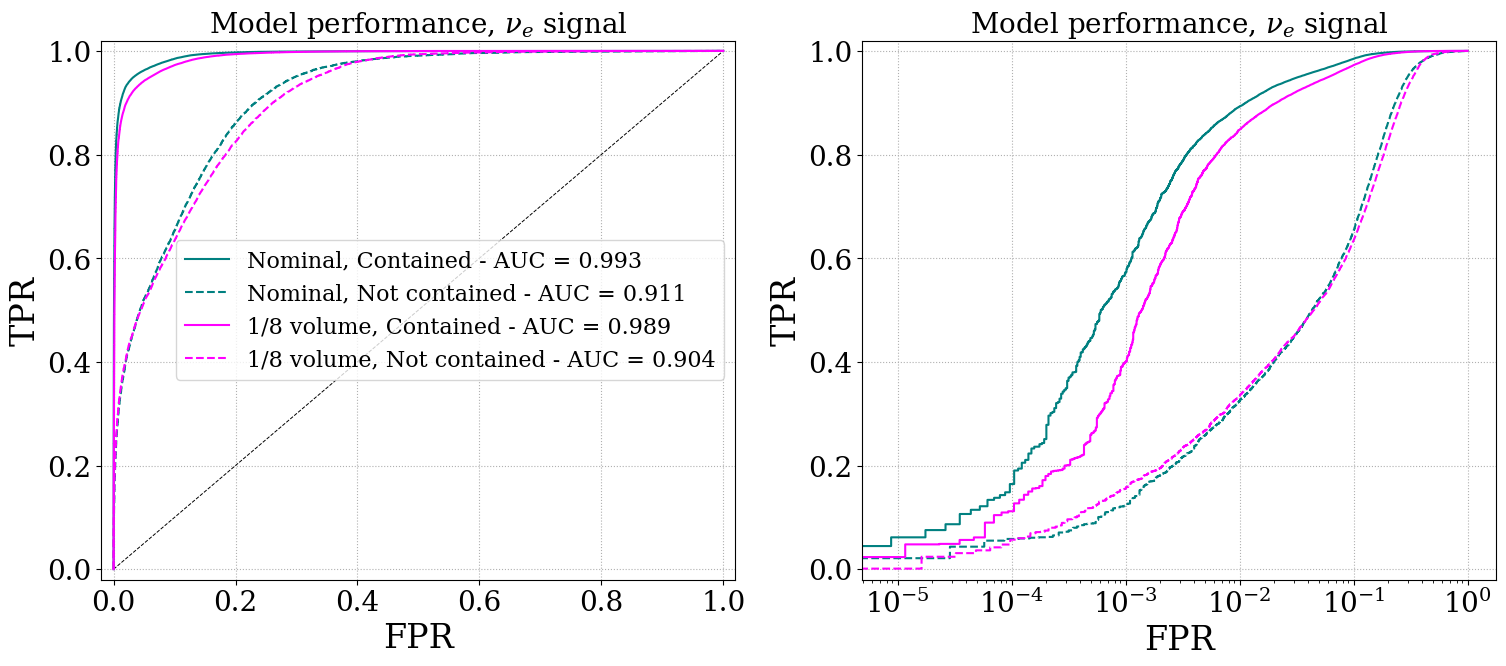}
    \caption{ROC curves for neutrino flavour classification models trained on the nominal and 1/8 volume datasets, using electron neutrino flavour as signal, split by containment of the produced charged lepton. The figure on the right is the same as on the left but with logarithmic x-axis to highlight the performance in the $10^{-3}$ FPR region, which is the target FPR for the electron neutrino sample. For both models, events where the charged lepton is contained in the detector volume have significantly better performance than events where it is not.}
    \label{fig:sizeSplitContainedsig}
\end{figure}
\begin{figure}[!t]
    \centering
    \includegraphics[width=1\linewidth]{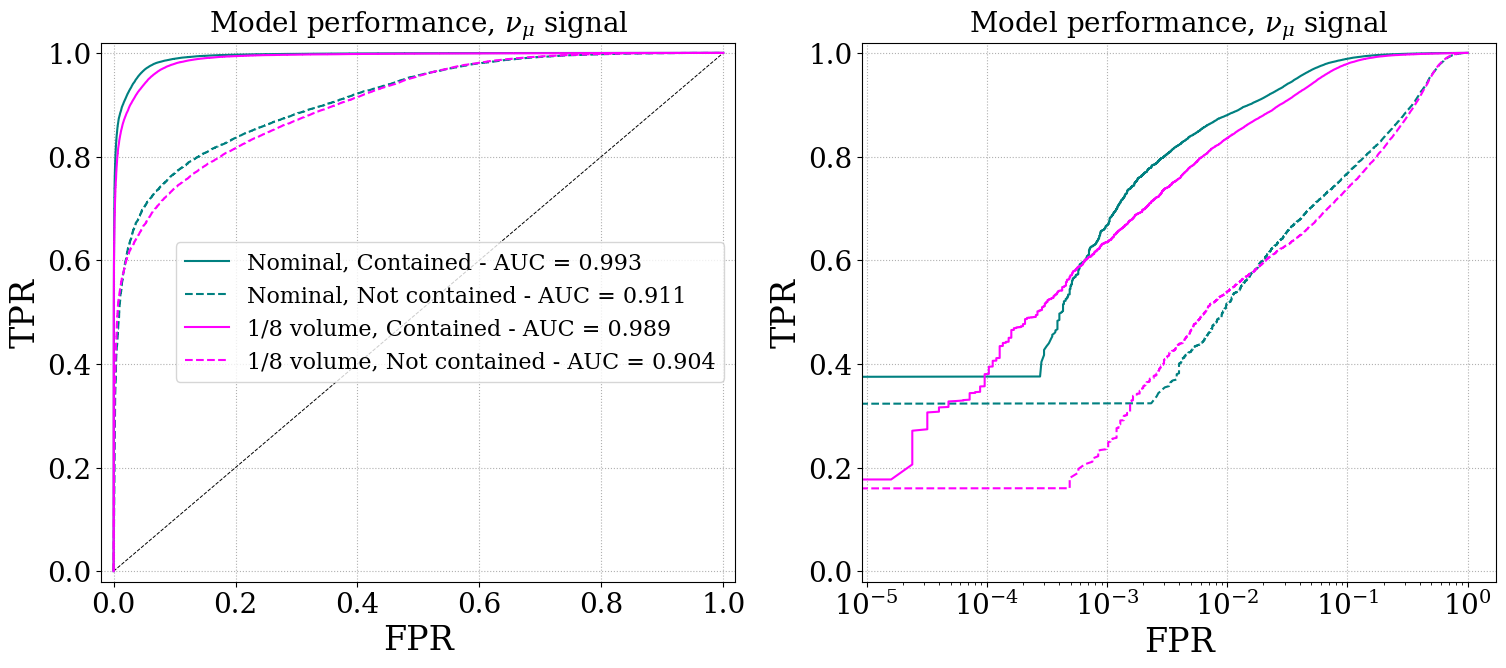}
    \caption{ROC curves for neutrino flavour classification models trained on the nominal and 1/8 volume datasets, using muon neutrino flavour as signal, split by containment of the produced charged lepton. The figure on the right is the same as on the left but with logarithmic x-axis to highlight the performance in the $10^{-2}$ FPR region, which is the target FPR for the muon neutrino sample. For both models, events where the charged lepton is contained in the detector volume have significantly better performance than events where it is not.}
    \label{fig:sizeSplitContainedbkg}
\end{figure}
To further test this hypothesis, the events were separated into events where the produced charged lepton comes to a stop and thus is contained within the detector volume and events where charged lepton leaves the detector volume. Figures \ref{fig:sizeSplitContainedsig} (electron flavour signal) and \ref{fig:sizeSplitContainedbkg} (muon flavour signal) show ROC curves for these two samples for the nominal and 1/8 volume detector sizes. For both sizes and both flavours, the events where the charged lepton is not contained have significantly worse performance, indicating that these are the events that can be rejected using the GNN to produce pure neutrino samples. ROC curves for all detector sizes are shown in Figures \ref{fig:sizeSplitContainedsigFull} and \ref{fig:sizeSplitContainedbkgFull} in Appendix \ref{sec:A-sizeSplitContained}.

\begin{figure}[!t]
    \centering
    \includegraphics[width=1\linewidth]{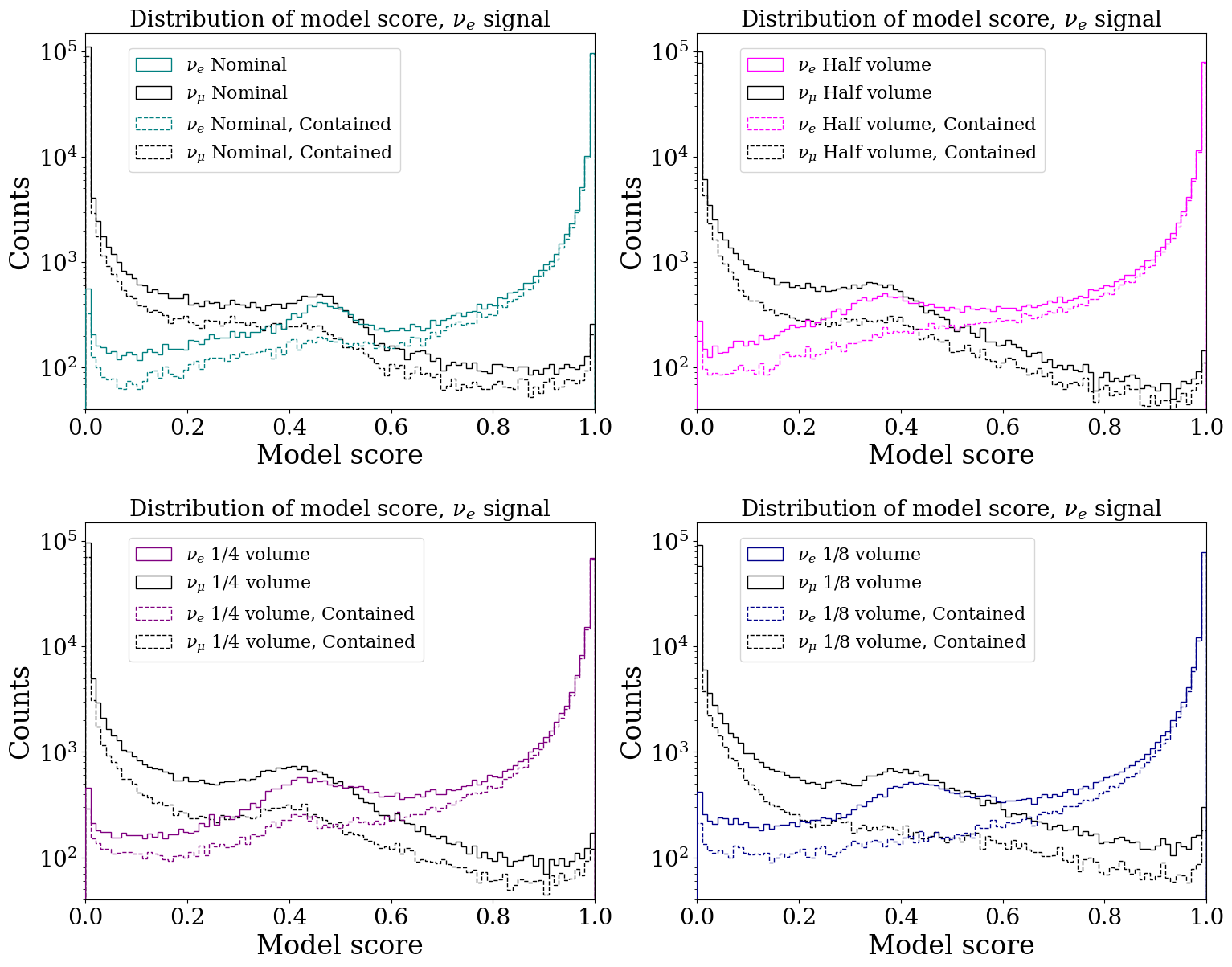}
    \caption{Model score distributions for neutrino flavour classification models trained on the nominal (top left), half volume (top right), 1/4 volume (bottom left) and 1/8 volume datasets (bottom right), for the full dataset (solid) and events with contained charged leptons (dashed). For all sizes, the full dataset exhibits a peak of ambiguous events near 0.4-0.6 that is not present among the contained events.}
    \label{fig:sizecompare_score}
\end{figure}

The model score distributions in Figure \ref{fig:sizecompare_score} confirm this analysis, showing the model score for the electron and muon neutrino samples for the full dataset and for the subset where the charged lepton is contained, for each detector size. This model score spectrum gives an indication of how the GNN models retain relatively high TPRs: The full event distribution contains a signifiant set of ambiguous events, which receive a model score close to 0.5 and are rejected from both samples rather than getting misidentified, manifesting as peak near 0.4-0.6. This peak is not equally prevalent in the contained sample, indicating that the model identifies the not contained events and assigns them an score associated with ambiguity. It is useful that in addition to the ability to reconstruct events with less PMT information, the GNN model through the model score also gives a natural estimate of the ambiguity of each event which is automatically considered when choosing threshold values. Figures \ref{fig:sizecompare_score_notContained} and \ref{fig:sizecompare_score_bothContained} in Appendix \ref{sec:A-sizeSplitContained} show the model score distributions for the full and not contained samples, and the contained and not contained samples. 

The performance for the different detector volume was also tested separately for neutrino and antineutrino events, the results of which are shown in Figures \ref{fig:sizeSplitPIDsig} (electron flavour signal) and \ref{fig:sizeSplitPIDbkg} (muon flavour signal) in Appendix \ref{sec:A-sizeSplitPID}. While the performance is very similar, the antineutrino events have consistently better performance for all sizes and flavours. The slight difference primarily arises from preference for forward scattering of the right-handed antineutrinos, which leads to lower momentum transfer and a stronger detector signal. This effect has also been observed in \cite{Alekou2022}.

Additionally, the energy dependency of the performance was tested by plotting the model score as a function of neutrino energy, for the the nominal and 1/8 volume detector sizes and for contained and not contained charged leptons, which is shown in Figure \ref{fig:sizeSplitContainedEnergy} in Appendix \ref{sec:A-splitcontainedenergy}. The model score exhibits an overall flat distribution for both sizes except for the 0-250 MeV range where the scores for both flavours are distributed around the 0.4-0.6 peak, as was observed in \cite{GNNKaare}. The low energy ambiguous events are not present in the not contained sample, as these events will almost never exit the detector. 

\subsection{Coverage Variation}

\subsubsection{Simulated Charge Recordings - Predicting PMT Importance}

\begin{figure}[!t]
    \centering
    \includegraphics[width=\linewidth]{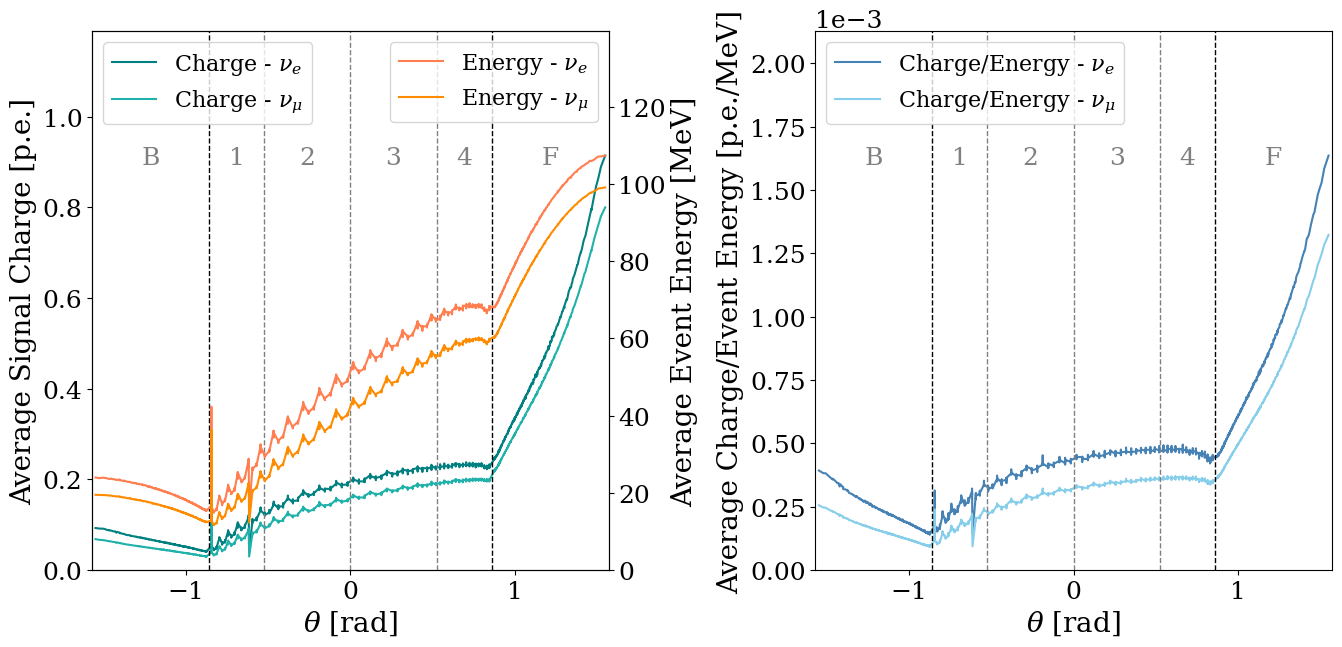}
    \caption{\textbf{Left:} The average charge per electron (muon) neutrino event recorded by a single PMT in green (cyan) and the average neutrino energies of signal hits in individual PMTs in red (orange) as a function of the angle with the beam direction $\theta$, averaged over all PMTs that share the same $\theta$ value. \textbf{Right:} The average recorded charge divided by the neutrino energy as a function of $\theta$.}
    \label{fig:PMTimportance_collapsed_sum}
\end{figure}

\begin{figure}[!t]
    \centering
    \includegraphics[width=1\linewidth]{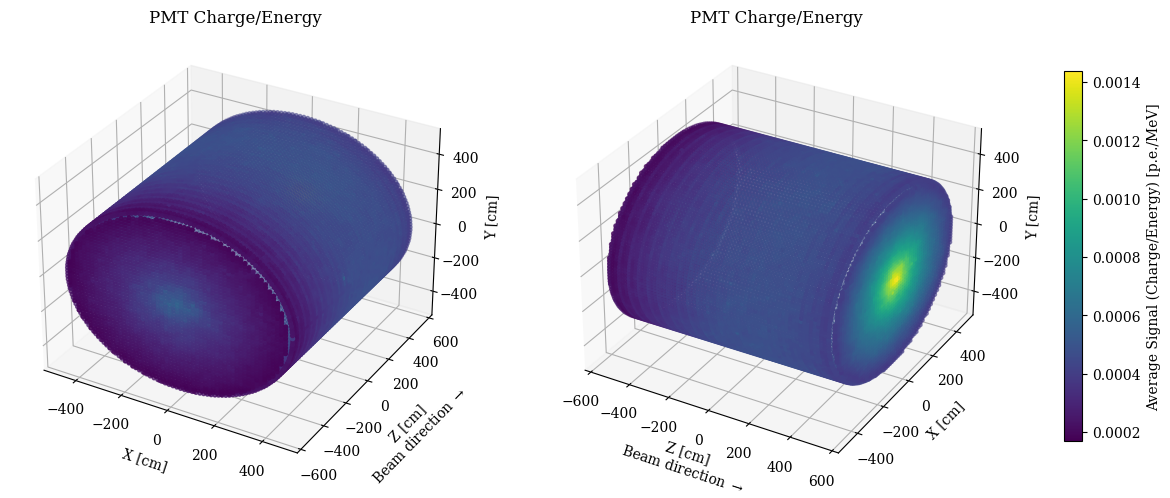}
    \caption{The average recorded charge per muon neutrino event divided by the neutrino energy for each PMT of the detector, showing the backward end-cap (left) and the forward end-cap (right). Generally, PMTs closer to the forward end-cap seem to record more charge per neutrino energy, with the most central PMTs at each end-cap recording the most charge per energy.}
    \label{fig:PMTimportance_3D}
\end{figure}

Before making changes to the PMT layouts, the simulated events and PMT signals themselves were studied to gain insight on factors that could be correlated with classification performance. Figure \ref{fig:PMTimportance_collapsed_sum} (left) shows the average charge per electron (muon) neutrino event recorded by a single PMT as a function of the position of the PMT represented by $\theta$, defined as the angle between the beam axis and the PMT position vector, measured from the centre of the detector (averaged over all PMTs that share the same $\theta$ value) in green (cyan), and the cumulative neutrino energy of the event for each hit for the same PMTs in red (orange). While the total charge recorded is expected to be correlated with the PMTs importance for classification, the total energy may be misleading, as events with higher energy, which are often easier to reconstruct, will contribute more to this value. The right side of Figure \ref{fig:PMTimportance_collapsed_sum}, shows the average recorded charge divided by the event energy for each PMT hits, which should give an indication of the ability of a PMT in each position to translate neutrino energy to recorded charge. Figure \ref{fig:PMTimportance_3D} shows the average charge over energy for muon neutrino events mapped onto the 3D surface of the detector. As expected, the PMTs in the forward end cap (in the beam direction), in particular central ones, record more charge, and PMTs in the forward part of the barrel do so as well to a lesser degree, with a slight dip in the corners. The PMTs at the backward end record less signal, with the exception of central end cap PMTs. 

This study was also performed separately for neutrinos and antineutrinos, and the results are shown in Figure \ref{fig:PMTsplit} in Appendix \ref{sec:A-PMTimportanceSplit}. For both muon and electron neutrinos, the PMTs collect more signal from antineutrino events than neutrinos, providing a possible explanation for the slightly better classification performance for antineutrinos over neutrinos found in Section \ref{sec:detectorsize}. The trend of PMTs towards the far end of the detector recording more signal is also enhanced for antineutrinos.

\subsubsection{Classification Performance}

\begin{figure}[!t]
    \centering
    \includegraphics[width=\linewidth]{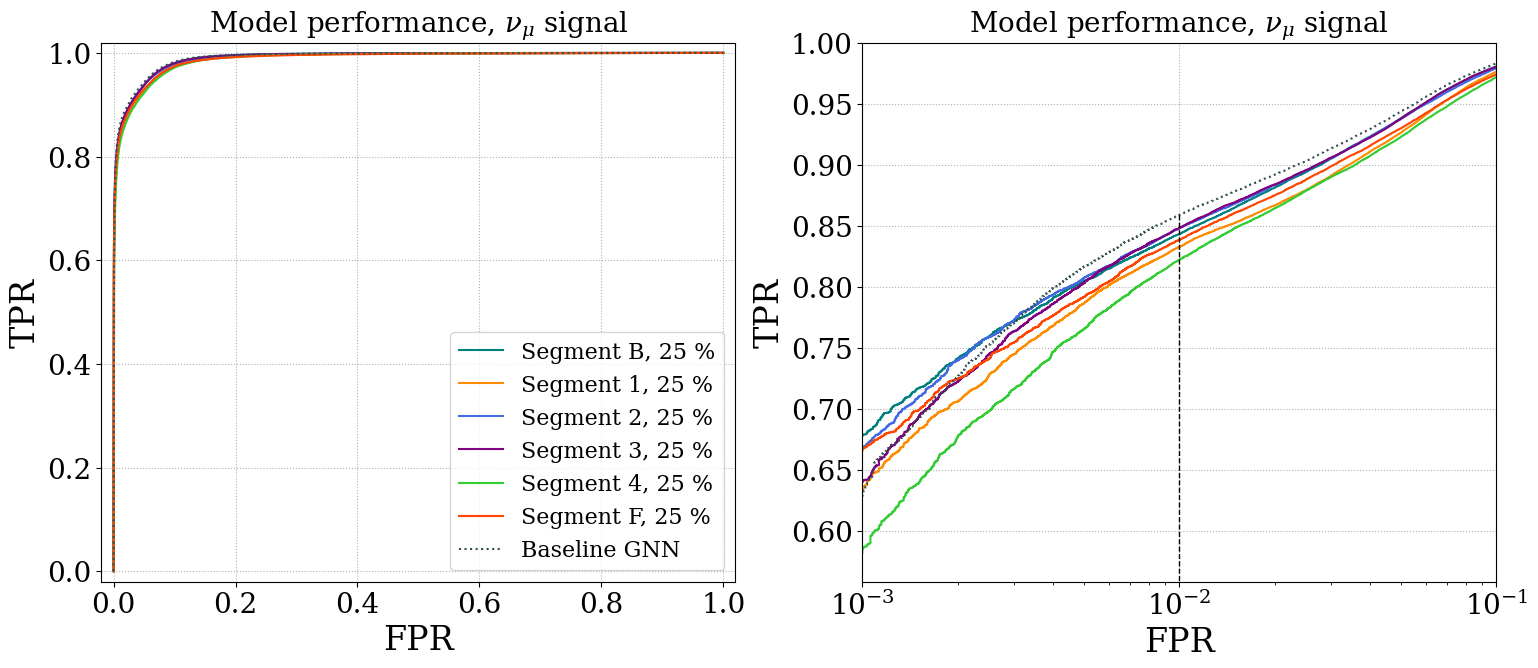}
    \caption{ROC curves for 6 GNN models trained on datasets with each of the 6 detector segments adjusted by ignoring 75 \% of the PMTs, and the baseline model trained on the full dataset. On the right, a dashed line indicates the target FPR for muon neutrinos of $10^{-2}$ and x-axis is logarithmic to show more detail near this value. At $10^{-2}$, the difference between the models is small, with resulting TPRs, which are equivalent to signal efficiency, ranging from around 0.82 to 0.85, which will not have a significant impact on the final analysis.}
    \label{fig:25off_bkg}
\end{figure}

To study the effect of varying the coverage across the detector, the 6 segments of the detector where adjusted individually. For each segment, a new model was trained on a dataset where signals from 75 \% of the PMTs in that segment were ignored according to the scheme described in Section \ref{sec:coverage}. Figure \ref{fig:25off_bkg} shows ROC curves for each of these 6 models and a baseline model trained on the full events. The full scale ROC curve on the left shows that the performance of the models is very similar, and the AUCs are very similar. The plot on the right has a log scale on the x-axis, and is zoomed in on the region where the target muon neutrino FPR of 0.01 is achieved, and shows more clearly the difference between the models. The target FPR is indicated with a vertical dashed line, and the resulting TPRs are shown. The baseline model as expected gives the highest TPR, while the reduced coverage of barrel segment 4, closest to the end-cap has the biggest impact on performance. 

\begin{figure}[!t]
    \centering
    \includegraphics[width=\linewidth]{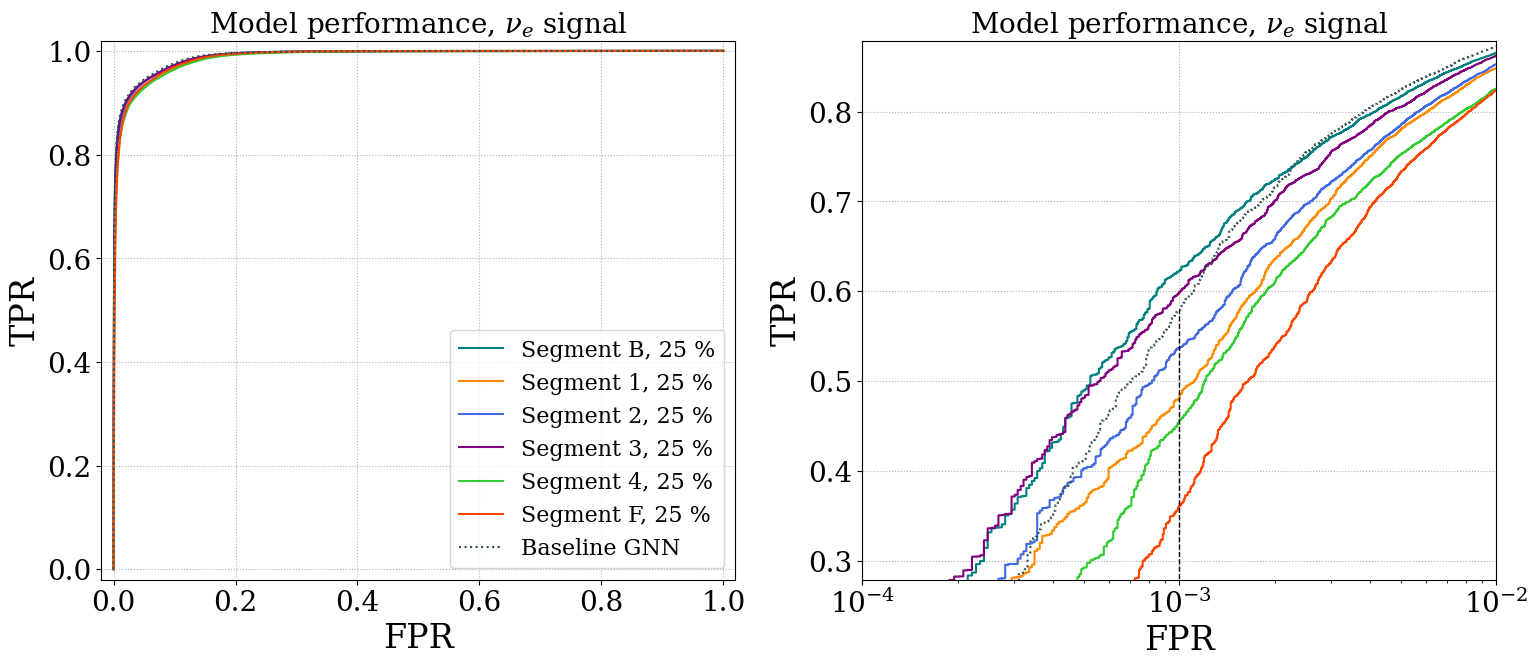}
    \caption{ROC curves for 6 GNN models trained on datasets with each of the 6 detector segments adjusted by ignoring 75 \% of the PMTs, and the baseline model trained on the full dataset. On the right, a dashed line indicates the target FPR for electron neutrinos and x-axis is logarithmic to show more detail near this value. At the lower FPR of $10^{-3}$, the resulting TPRs vary significantly from below 0.4 to higher than 0.6, which will result in a more significant impact on the physics analysis.}
    \label{fig:25off_sig}
\end{figure}

Figure \ref{fig:25off_sig} shows the same ROC curves as Figure \ref{fig:25off_bkg}, but using electron neutrinos as the signal events. Here, the target FPR is 0.001, which is indicated on the right. This plot shows that the GNN performance is still quite similar for all models, with coverage reduction in the forward end-cap resulting in the greatest decrease in FPR. Table \ref{tab:AUCTPR} shows the resulting TPRs for both the muon and electron samples, as well as the AUCs for all the segments and the difference from the baseline for each of these parameters. Similarly to the results in Section \ref{sec:detectorsize}, the GNN models obtain a reasonable performance for all of the detector configurations, as the resulting TPRs are within the same order of magnitude, between 0.360 and 0.622 (0.822 and 0.859) with electron (muon) neutrino target, and any decrease in efficiency from the baseline can be converted directly to increased runtime of the detector. This again illustrates the trade-off between instrumentation and runtime costs to obtain the same number of events for the final physics analysis. For segments B and 3, the coverage reductions result in slightly higher TPRs for electron neutrinos, but since the increase is only by 0.043 and  0.019, and the AUCs and muon neutrino TPRs still decrease, the collective performance on both flavours (which the model is optimized for) is still reduced. Generally, there is a hierarchy in this coarse segmentation of PMTs, with adjustments of the coverage of the forward end cap (F) as well as Segments 4 and 1 having the most significant negative impact on classification performance with TPR respective reductions of 0.219, 0.125, and 0.099 (0.021, 0.037, and 0.026) for electron (muon) neutrino signal. 

\begin{table}
    \centering
    \begin{tabular}{ccccccc}
         \textbf{Adjusted segment} & \textbf{AUC} & \textbf{$\Delta$ AUC} & \textbf{TPR ($\nu_\mu$)} & \textbf{$\Delta$TPR ($\nu_\mu$)} & \textbf{TPR ($\nu_e$)} & \textbf{$\Delta$TPR ($\nu_e$)} \\
         Baseline & 0.991 & - & 0.859 & - & 0.579 & - \\
         \hline
         Segment B & 0.990 & -0.001 & 0.843 & -0.016 & 0.622 & +0.043 \\
         Segment 1 & 0.989 & -0.002 & 0.833 & -0.026 & 0.480 & -0.099 \\
         Segment 2 & 0.990 & -0.001 & 0.848 & -0.011 & 0.537 & -0.042 \\
         Segment 3 & 0.990 & -0.001 & 0.848 & -0.011 & 0.598 & +0.019 \\
         Segment 4 & 0.987 & -0.004 & 0.822 & -0.037 & 0.454 & -0.125 \\
         Segment F & 0.988 & -0.003 & 0.838 & -0.021 & 0.360 & -0.219 \\
    \end{tabular}
    \caption{The obtained summary Area Under Curve (AUC) for the 6 models each trained with one of detector segments adjusted to 25 \% PMT coverage, and the resulting True Positive Rates (TPRs) obtained when applying the selection criteria for muon and electron neutrinos, respectively. Also shown are the differences in AUC, and TPRs between each model and the baseline. The segments for which reducing the coverage has the greatest impact on performance are Segment F, Segment 4, and Segment 1, while reducing the coverage of the backwards end-cap Segment B and the central barrel segments 2 and 3 has a comparably small impact on performance.}
    \label{tab:AUCTPR}
\end{table}

\section{Conclusion}

This work explores the use of graph neural network-based classification to guide the design of the ESS$\nu$SB near water Cherenkov detector. By systematically varying detector size and PMT coverage in event simulation and training GNN models for each configuration, we have quantified how detector performance depends on key design parameters.

We have demonstrated how GNN-based classification remains effective even for substantially reduced detector volumes. While classification performance decreases with decreasing size, the degradation is moderate and primarily manifests as a reduction in signal efficiency at fixed background rejection. Importantly, the models retain the ability to separate electron and muon neutrino interactions even for detectors with volumes reduced by almost an order of magnitude. This indicates that reduced detector size can be traded against increased runtime, providing a clear cost-performance lever for detector design that is particularly attractive for a near detector that will have high interaction rates.

The study of non-uniform PMT coverage shows that classification performance is relatively insensitive to coverage reductions in large parts of the detector. However, a clear hierarchy emerges: coverage in the forward end-cap and adjacent barrel regions has the largest impact, while reductions in the backward end-cap and central barrel regions are less critical. This is consistent with the distribution of detected charge, which is highest in the forward direction of the neutrino beam. These results suggest that optimized, non-uniform instrumentation schemes can be considered without significant loss of performance.

In addition to measuring the composition of the neutrino beam, the ESS$\nu$SB Near Detector will also monitor the neutrino energy spectrum, which is inferred from reconstructed charged lepton energy and direction. These variables can be reconstructed using similar GNN models. Although flavour identification and regression tasks depend differently on the detector response and the underlying algorithms, the finding that GNN flavour classification can be performed with reduced detector sizes indicates that robust event-level features can still be extracted. Future studies will establish if these techniques can be extended to regression tasks while maintaining acceptable resolution. 

Overall, this work demonstrates that GNN-based classification provides both strong performance and significant flexibility in detector design. The ability to maintain robust classification under reduced size and non-uniform coverage opens the possibility of optimizing the ESS$\nu$SB near detector for cost and feasibility while preserving sensitivity to the underlying physics.

\newpage
\appendix

\newpage
\section{Detector size variation performance separated by containment of charged lepton}\label{sec:A-sizeSplitContained}

\begin{figure}[!ht]
    \centering
    \includegraphics[width=\linewidth]{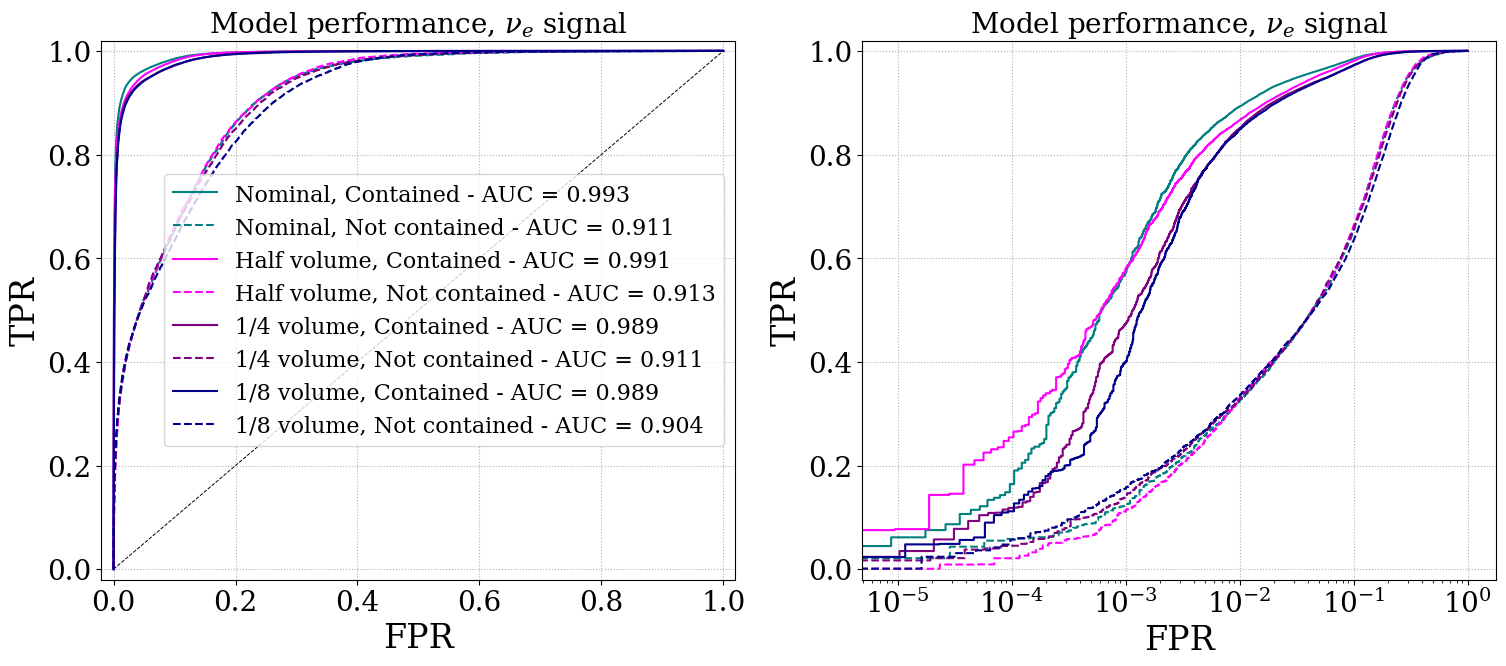}
    \caption{ROC curves for neutrino flavour classification models trained on the nominal, half volume, 1/4 volume, and 1/8 volume datasets, using electron neutrino flavour as signal, split by containment of the produced charged lepton. The figure on the right is the same as on the left but with logarithmic x-axis to highlight the performance in the $10^{-3}$ FPR region. Events where the charged lepton is contained in the detector volume have significantly better performance.}
    \label{fig:sizeSplitContainedsigFull}
\end{figure}

\begin{figure}[!ht]
    \centering
    \includegraphics[width=\linewidth]{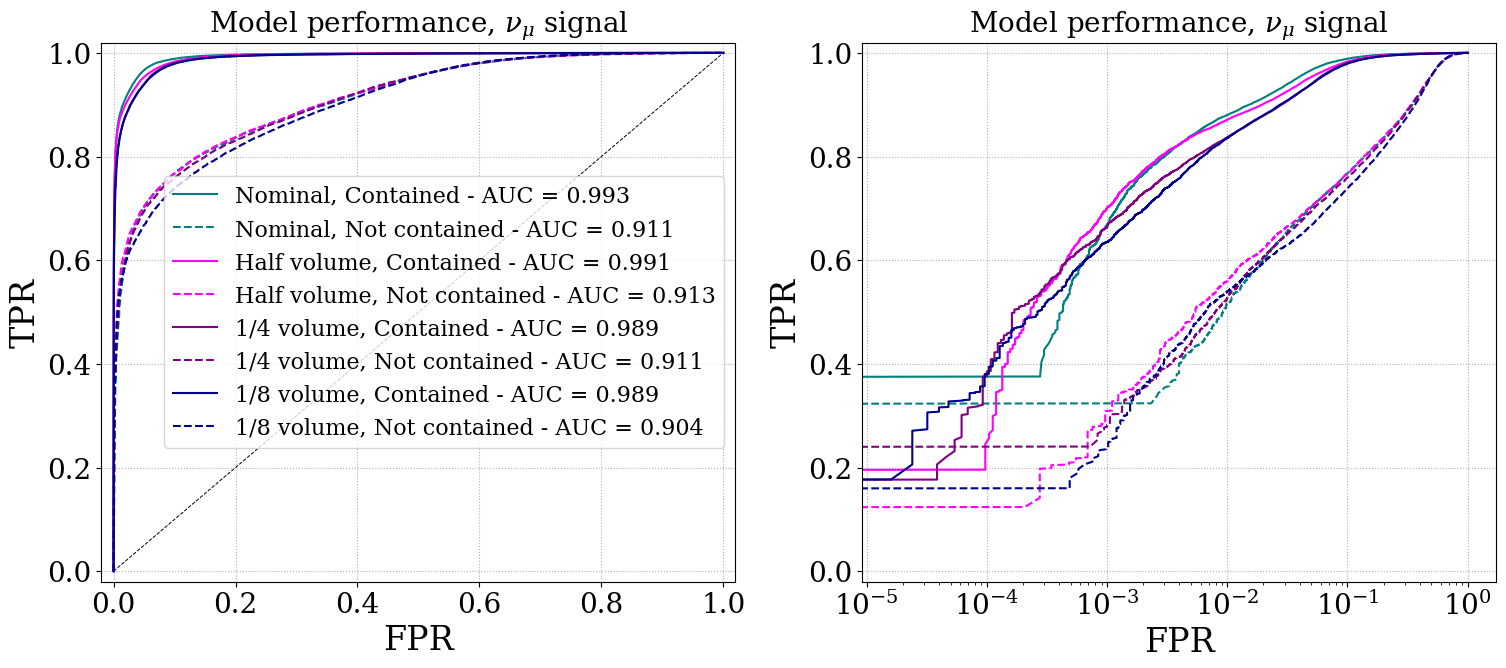}
    \caption{ROC curves for neutrino flavour classification models trained on the nominal, half volume, 1/4 volume, and 1/8 volume datasets, using muon neutrino flavour as signal, split by containment of the produced charged lepton. The figure on the right is the same as on the left but with logarithmic x-axis to highlight the performance in the $10^{-2}$ FPR region. Events where the charged lepton is contained in the detector volume have significantly better performance.}
    \label{fig:sizeSplitContainedbkgFull}
\end{figure}

\newpage
\begin{figure}[!ht]
    \centering
    \includegraphics[width=1\linewidth]{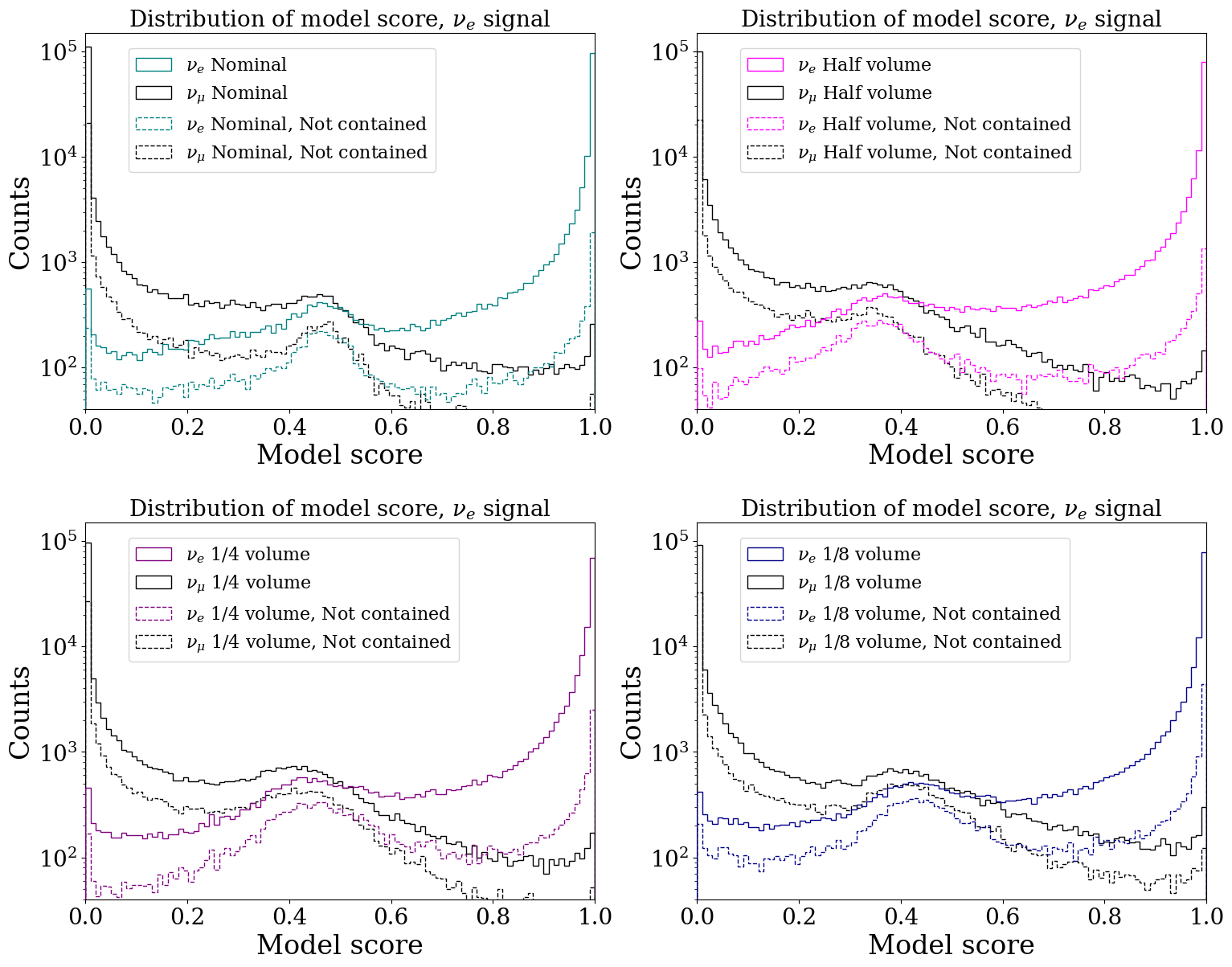}
    \caption{Model score distributions for neutrino flavour classification models trained on the nominal (top left), half volume (top right), 1/4 volume (bottom left) and 1/8 volume datasets (bottom right), for the full dataset (solid) and events with not contained charged leptons (dashed). For all sizes, the full dataset exhibits a peak of ambiguous events near 0.4-0.6 that is even more pronounced among the not contained events.}
    \label{fig:sizecompare_score_notContained}
\end{figure}

\newpage
\begin{figure}[!ht]
    \centering
    \includegraphics[width=1\linewidth]{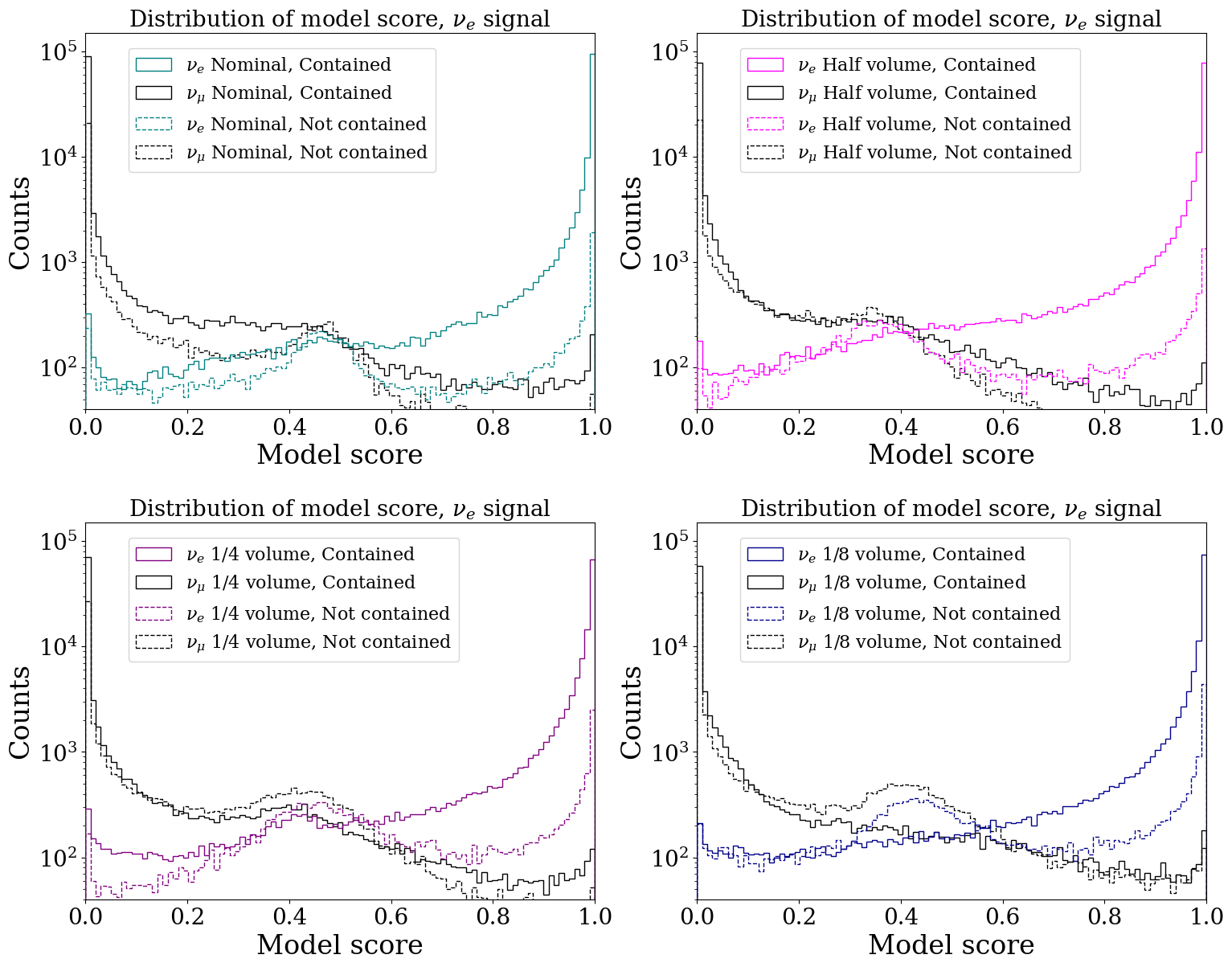}
    \caption{Model score distributions for neutrino flavour classification models trained on the nominal (top left), half volume (top right), 1/4 volume (bottom left) and 1/8 volume datasets (bottom right), for the full dataset (solid) and events with contained charged leptons (dashed). For all sizes, the not contained dataset exhibits a peak of ambiguous events near 0.4-0.6 that is not present among the contained events.}
    \label{fig:sizecompare_score_bothContained}
\end{figure}

\newpage
\section{Detector size variation performance separated by $\nu$/$\bar{\nu}$}\label{sec:A-sizeSplitPID}

\begin{figure}[!ht]
    \centering
    \includegraphics[width=\linewidth]{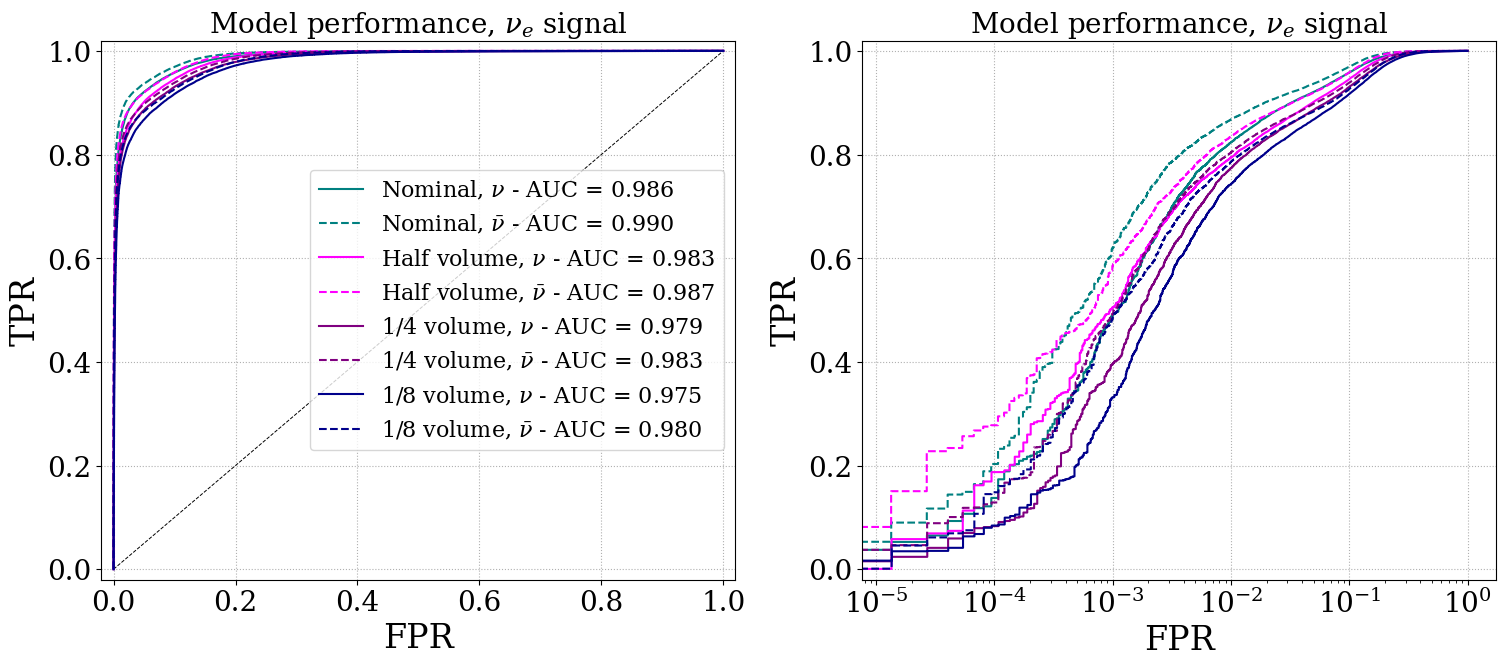}
    \caption{ROC curves for neutrino flavour classification models trained on the nominal, half volume, 1/4 volume, and 1/8 volume datasets, using electron neutrino flavour as signal, split by neutrino and antineutrino events. The figure on the right is the same as on the left but with logarithmic x-axis to highlight the performance in the $10^{-3}$ FPR region. Antineutrino events have slightly better classification performance for all detector sizes then neutrino events.}
    \label{fig:sizeSplitPIDsig}
\end{figure}

\begin{figure}[!ht]
    \centering
    \includegraphics[width=\linewidth]{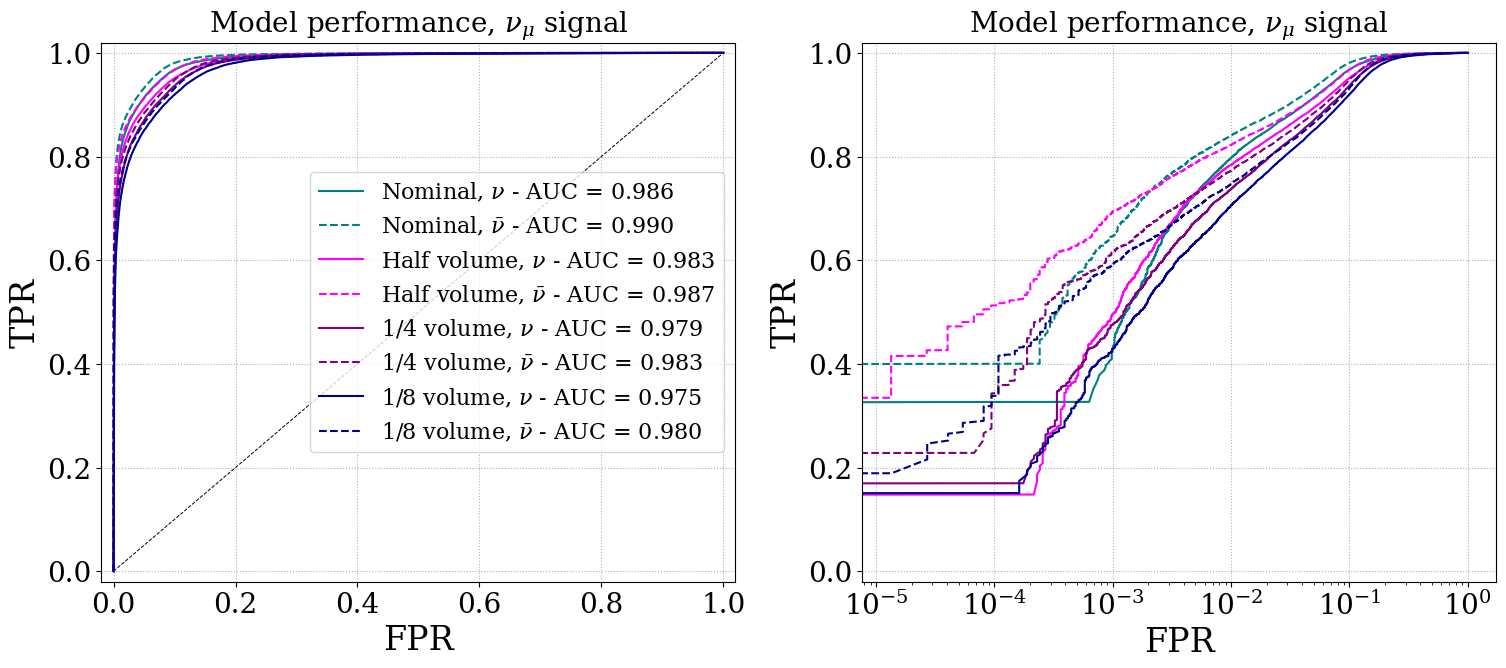}
    \caption{ROC curves for neutrino flavour classification models trained on the nominal, half volume, 1/4 volume, and 1/8 volume datasets, using muon neutrino flavour as signal, split by neutrino and antineutrino events. The figure on the right is the same as on the left but with logarithmic x-axis to highlight the performance in the $10^{-2}$ FPR region. Antineutrino events have slightly better classification performance for all detector sizes then neutrino events.}
    \label{fig:sizeSplitPIDbkg}
\end{figure}

\newpage
\section{Model score as a function of neutrino energy separated by containment of charged lepton}\label{sec:A-splitcontainedenergy}

\begin{figure}[!ht]
    \centering
    \includegraphics[width=\linewidth]{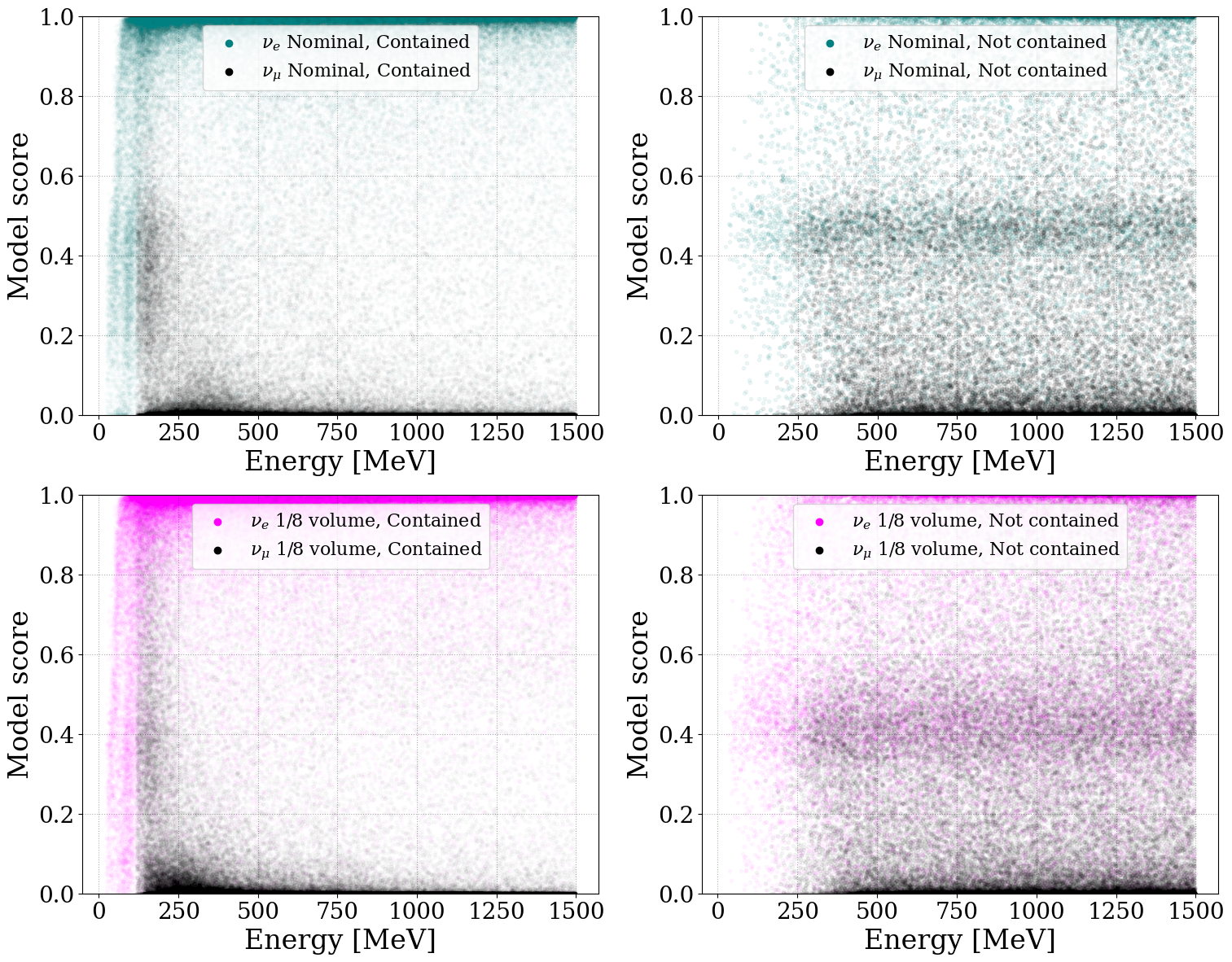}
    \caption{Model score for muon and neutrino events as a function of the neutrino energy for the nominal (top) and 1/8 volume (bottom) datasets, and for contained (left) and not contained (right) events. Low energy events with contained charged leptons receive model scores closer to 0.5, indicating ambiguous events.}
    \label{fig:sizeSplitContainedEnergy}
\end{figure}

\newpage
\section{PMT charge recordings separated by $\nu$/$\bar{\nu}$}\label{sec:A-PMTimportanceSplit}

\begin{figure}[!ht]
    \centering
    \includegraphics[width=\linewidth]{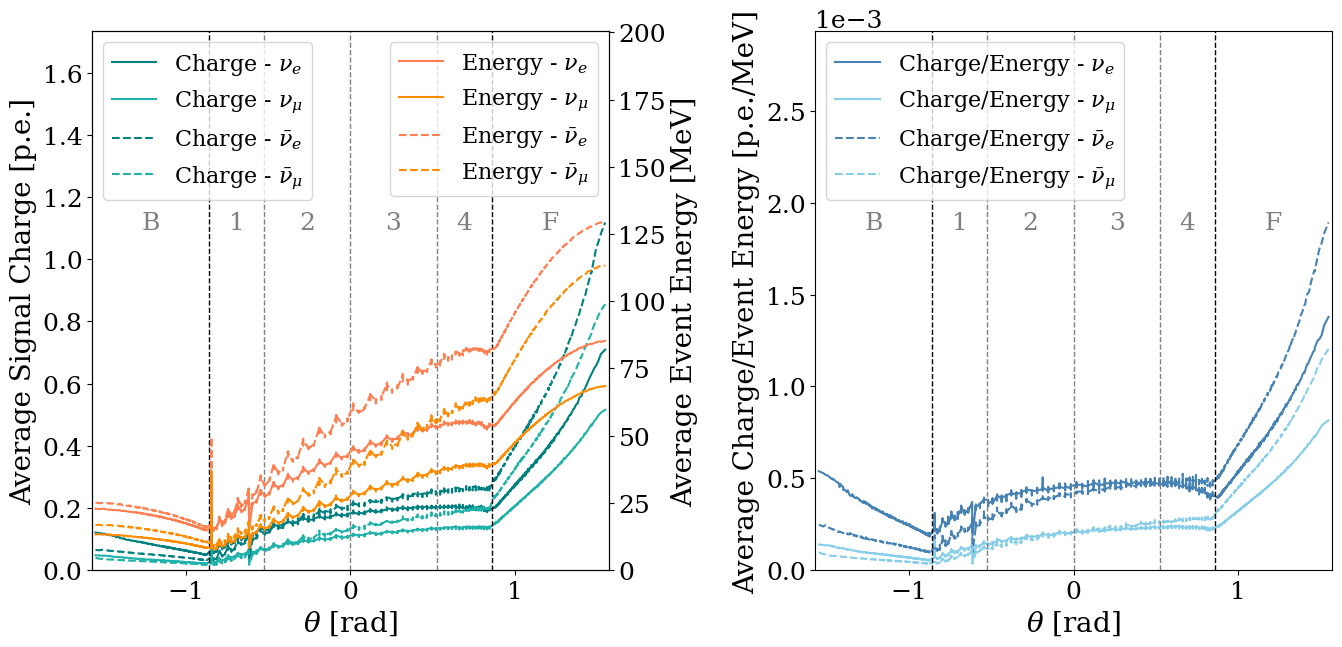}
    \caption{\textbf{Left:} The average charge per electron (muon) neutrino event recorded by a single PMT in green (cyan) and the average neutrino energies of signal hits in individual PMTs in red (orange) as a function of the angle with the beam direction $\theta$, averaged over all PMTs that share the same $\theta$ value. \textbf{Right:} The average recorded charge divided by the neutrino energy as a function of $\theta$. In both figures, the events are separated in neutrino (solid) and antineutrino (dashed) events.}
    \label{fig:PMTsplit}
\end{figure}

\acknowledgments

Funded by the European Union. Views and opinions expressed are however those of the author(s) only and do not necessarily reflect those of the European Union. Neither the European Union nor the granting authority can be held responsible for them.

The authors acknowledge support provided by the European Union’s Horizon 2020 research and innovation programme under the Marie Skłodowska -Curie grant agreement No 860881-HIDDeN, and the Croatian Science Foundation under the project number HRZZ-DOK-NPOO-2023-10-1262; Swiss National Science Foundation and Croatian Science Foundation via grant MAPS IZ11Z0-230193; Ministry of Science, Education and Youth of Republic of Croatia via grant No. PK.1.1.10.0002.

The training of models and classification of events were performed using the GraphNeT framework (Apache 2.0), developed and maintained by GraphNeT group. The computations were enabled by resources provided by the National Academic Infrastructure for Supercomputing in Sweden (NAISS) and the Swedish National Infrastructure for Computing (SNIC) at LUNARC partially funded by the Swedish Research Council through grant agreements no. 2022-06725 and no. 2018-05973. 

The authors acknowledge the assistance from C. Vilela, E. O’Sullivan, H. Tanaka, B. Quilain and M. Wilking for the use of the WCSim and fiTQun software packages.






\bibliographystyle{JHEP}
\bibliography{biblio.bib}



\end{document}